\documentclass[11pt,oneside, a4paper]{article}
\pdfoutput=1

\ifx\pdfoutput\undefined
\usepackage[dvips,bookmarks=false]{hyperref}	% This is for arXiv.org
\else
\usepackage{hyperref}	% This is for pdftex
\fi
\hypersetup{colorlinks,bookmarksopen,bookmarksnumbered,citecolor=blue,
linkcolor=black,pdfstartview=FitH,urlcolor=blue}

\def\mhref#1{\href{mailto:#1}{#1}}		% email on title page

\newcommand{\1}{\mbox{1}\hspace{-0.25em}\mbox{l}}
\usepackage{slashed}
\usepackage{graphicx}
\usepackage{amssymb}
\usepackage{cite}
\usepackage{bm}
\usepackage{indentfirst}
\usepackage{amsmath}
\usepackage{hhline}
\usepackage{multirow}
\usepackage{soul}
\usepackage{float}
\def\thefootnote{\fnsymbol{footnote}} % use this to have fancy footnote symbols, remove for numbers

\newcommand{\<}{\langle}
\renewcommand{\>}{\rangle}
\def\be{\begin{equation}}
\def\ee{\end{equation}}
\newcommand{\bea}{\begin{eqnarray}}
\newcommand{\eea}{\end{eqnarray}}

\newcommand{\bee}[0]{\begin{eqnarray}}
\newcommand{\eee}[0]{\end{eqnarray}}

\usepackage[textwidth=1.7cm,textsize=tiny]{todonotes}

\allowdisplaybreaks

\begin{document}
\begin{titlepage}

\begin{flushright}
 LPT-Orsay-18-63
\end{flushright}

\begin{center}

\vspace{1cm}
{\Large\bf 
Beta and Neutrinoless Double Beta Decays  with KeV Sterile Fermions
}

\vspace*{0.8cm}

{\bf Asmaa Abada$^{a}$%
\footnote{\tt \mhref{asmaa.abada@th.u-psud.fr}}, 
\'Alvaro Hern\'andez-Cabezudo$^{b}$%
\footnote{\tt \mhref{alvaro.cabezudo@kit.edu}},\\ 
and 
Xabier Marcano$^{a}$%
\footnote{\tt \mhref{xabier.marcano@th.u-psud.fr}} }  
   
\vspace*{.5cm} 
$^{a}$Laboratoire de Physique Th\'eorique, CNRS, \\
Univ. Paris-Sud, Universit\'e Paris-Saclay, 91405 Orsay, France
\vspace*{.2cm} 

$^{b}$ Institut f\"ur Kernphysik, Karlsruher Institut f\"ur Technologie, 76021 Karlsruhe, Germany
\vspace*{.2cm}

\abstract{      
Motivated by the capability of the KATRIN experiment to explore the existence of KeV neutrinos in the $[1-18.5]$ KeV mass range, 
we explore the viability of minimal extensions of the Standard Model involving sterile neutrinos (namely the 3 + $N$ frameworks) and study their possible impact in both the beta energy spectrum and the neutrinoless double beta decay effective mass, for the two possible ordering  cases for the light neutrino spectrum. 
We also explore how both observables can discriminate between motivated low-scale seesaw realizations involving KeV sterile neutrinos. 
Our study concerns the prospect of a Type-I seesaw with two right-handed neutrinos, and a combination of the inverse and the linear seesaws where the Standard Model is minimally extended by two quasi-degenerate sterile fermions. 
We also discuss the possibility of exploring the latter case searching for double-kinks in KATRIN.}

\end{center}
\end{titlepage}

\renewcommand{\thefootnote}{\arabic{footnote}}
\setcounter{footnote}{0}

\setcounter{page}{2}

%%%%%%%%%%%%%%%%%%%%%%%%%%%%%%%%%%%%%%

\section{Introduction}
Neutrino oscillations constitute evidence for flavor violation in the neutral lepton sector suggesting the 
need to extend the Standard Model (SM) in order to account for the 
necessarily massive neutrinos and for the lepton mixing, as observed. The most minimal or trivial  extension of the SM is to consider the existence of right-handed (RH) neutrinos, producing thus the conventional Dirac mass terms for neutrinos. However the Majorana possible nature of the RH neutrinos is uncircumventable as is the question of the tiny active neutrino masses. The alternative solution is the embedding of the seesaw (Type-I) mechanism~\cite{Minkowski:1977sc,GellMann:1980vs,Yanagida:1979as,Glashow:1979nm,Mohapatra:1979ia,Schechter:1980gr,Schechter:1981cv} predicting Majorana nature for both the light active neutrinos and the heavy ones. One of the consequences is the violation of the total lepton number. 
Alternatively,  the observation of any lepton number violating (LNV) process will point towards the existence of New Physics (NP) and indirectly to the  Majorana nature of neutrinos.

Adding new neutral fermions to the SM field content leads to a broad range of new phenomenology: depending on their mass scale of these neutrinos, they may address open questions in astrophysics~\cite{Kusenko:2006rh,Kusenko:2008gh,Petraki:2007gq}, cosmology\footnote{A detailed review of the cosmological motivations for (light) sterile fermions can be found in~\cite{Abazajian:2012ys,Abazajian:2017tcc}.} (baryogengesis via leptogenesis, dark matter candidate, ...), or lead to interesting signals in laboratory experiments (beam-dump experiments, neutrinoless double beta decay, ...). 
In this study, we focus on minimal low-scale seesaw 
realizations~\cite{Gavela:2009cd,Ibarra:2010xw,Donini:2011jh,Abada:2014vea,Asaka:2005an} which can account for the observed neutrino masses and mixings. Note that this mechanism can also successfully generate  the Baryon Asymmetry of the Universe (BAU) via leptogenesis\footnote{The mechanism behind leptogenesis is the so-called ``ARS" mechanism,  first proposed by Akhmedov, Rubakov and Smirnov~\cite{Akhmedov:1998qx}, in which a lepton asymmetry is produced by the CP-violating oscillations of a pair of heavy sterile neutrinos.}, when the sterile neutrino masses are not exceeding about 50~GeV. This mass regime is also very interesting for LNV processes like the ``neutrinoless" meson and tau decay processes (see for instance~\cite{Atre:2009rg,Abada:2017jjx}) potentially giving rise to interesting collider signatures.

Interestingly, sterile neutrinos are present in several neutrino mass models and their existence is strongly motivated
by current  reactor  neutrino oscillation anomalies~\cite{Mueller:2011nm,Huber:2011wv,Mention:2011rk},  suggesting that there might be some extra fermionic gauge singlet(s) with mass(es) in the eV range \cite{Dentler:2018sju}. Their existence  is also motivated by  indications from large scale structure formation~\cite{Kusenko:2009up,Abazajian:2012ys}. Moreover, depending on their mass scale, sterile fermion states can also give rise to interesting collider signatures~\cite{Atre:2009rg,Dev:2013wba,Adams:2013qkq,Bonivento:2013jag,Blondel:2013ia,Das:2014jxa,Das:2015toa,Deppisch:2015qwa,Alekhin:2015byh,Anelli:2015pba,Banerjee:2015gca,Abada:2017jjx,Alva:2014gxa,Arganda:2015ija,Degrande:2016aje,Ruiz:2017yyf,Cai:2017mow,Pascoli:2018rsg}.

 On the other hand, models with sterile fermions are severely constrained\footnote{Due  to the presence of the additional sterile fermionic states, 
the modified neutral and charged lepton currents 
might lead to new contributions to a
vast array of observables, possibly in conflict with current
bounds and the SM extensions via sterile fermions 
must then be confronted to all available constraints
arising from high-intensity, high-energy and cosmological
observations. 
} from electroweak (EW) precision observables, laboratory data and cosmology, due to the mixings of the sterile  states with the active left-handed neutrinos. All the constraints that we take into account in our study are discussed in Appendix \ref{sec:constraints}. 

In this study we will be interested in sterile neutrinos in the KeV regime, which can impact the electron energy spectrum in tritium $\beta$ decays~\cite{Shrock:1980vy}. 
Indeed, following this idea, the KATRIN experiment~\cite{Osipowicz:2001sq,Angrik:2005ep}  could be able to probe KeV~\cite{Mertens:2014nha,Mertens:2014osa,Mertens:2015ila,Steinbrink:2017ung}, and also eV~\cite{Riis:2010zm,Formaggio:2011jg,Esmaili:2012vg}, sterile neutrinos with an unprecedented sensitivity in other laboratory experiments\footnote{ After the KATRIN direct neutrino mass measurement  program is completed, TRISTAN will prospectively be integrated in the beam-line of KATRIN experiment in 2025~\cite{Boyarsky:2018tvu}, with the aim of measuring the full $\beta$ spectrum. TRISTAN is being previously implemented in the Troitsk nu-mass experiment.}. 
Interestingly,  the neutrinoless double beta decay ($0\nu 2 \beta$) - which is  by excellence the observable associated with the existence of Majorana neutrinos - when mediated by sterile neutrinos appears to be the ideal laboratory to 
probe their parameter space as the $0\nu 2 \beta$ amplitude is affected by their presence.

However, in the SM extensions we consider in this work, where the  mass(es) and active-sterile mixings of the sterile states are within the KATRIN (TRISTAN) experiment sensitivity reach, we expect that the cosmological constraints (from Big Bang Nucleosynthesis and $N_{\rm{eff}}$) to be particularly severe, see~\cite{Hernandez:2014fha}. Most importantly, the  astrophysical bounds are currently  some orders of magnitudes stronger
than any laboratory limit, as  sterile neutrinos can  decay~\cite{Shrock:1974nd} and X-ray observations provide
bounds on their parameters, see for instance~\cite{Loewenstein:2008yi,Loewenstein:2012px}. 
Nevertheless, these limits rely on underlying
cosmological and astrophysical assumptions and can be evaded as argued  in for instance~\cite{Gelmini:2004ah,Gelmini:2008fq,Bezrukov:2017ike}. 
Consequently, it is important to perform laboratory searches that could provide independent and complementary information to that from cosmological and astrophysical observations.
A detection through production of a KeV sterile neutrino in KATRIN for instance, would be completely independent of any cosmological
and  astrophysical input and has the potential to independently test  the sterile  hypothesis.

Thus our study concerns laboratory probes of  extensions of the SM with sterile fermions with mass ranges leading to possible impact in the KATRIN energy spectrum and also in the neutrinoless double beta decay effective mass, which (adapted) expression is given in~\cite{Benes:2005hn,Blennow:2010th}.  
 In order to understand and illustrate the impact due to the presence of sterile fermions on the latter observables, the neutrino effective Majorana mass $m_{ee}$ in $0\nu 2 \beta$ and the $\beta$ decay neutrino effective mass, $m_\beta$, we start by considering a bottom-up approach, which consist in adding to the SM a certain number $N$ of sterile neutrinos, making no hypothesis on the neutrino mass generation mechanism, in order to capture some of the effects of more complicated frameworks (like the seesaw mechanisms we consider in this study). 
This will provide a useful first approach before we consider 
explicit minimal seesaw models capable of accommodating neutrino data. The seesaw models we consider necessitate the introduction of neutral fermion fields belonging to two categories: (i) RH neutrinos, which in the interaction basis feature Yukawa interactions with the SM Higgs and lepton doublets, namely the Type-I seesaw at a low enough Majorana mass scale (typically with small Yukawa couplings),  and (ii) sterile neutrinos, which have no such couplings. In a slight abuse of notation, we will also apply this classification to the Linear Seesaw Mechanism (LSS)~\cite{Barr:2003nn,Malinsky:2005bi} and to the Inverse Seesaw mechanism (ISS)~\cite{Wyler:1982dd,Mohapatra:1986bd,GonzalezGarcia:1988rw}, in which cases the `sterile' neutrinos in fact have (very suppressed) couplings to the SM neutrinos. Most of our analysis will be however carried out in the mass basis, where the new states are in general a mixture of the RH and sterile (and active) components. We will thus more generally refer to states dominated by RH and/or sterile components as (SM) fermionic singlets.
We will be particularly interested in addressing minimalistic realizations of low-scale seesaw mechanisms that are the Type-I with two RH neutrinos, as well as,   a combination of a linear and an inverse seesaw involving two sterile neutrinos (which we will name from now on ``LISS"\footnote{This minimal model was used in~\cite{Abada:2017ieq} and named LSS-ISS.}).  
This kind of signatures have also been explored in other contexts, such as Left-Right symmetric models~\cite{Barry:2014ika}, extra dimensions~\cite{Rodejohann:2014eka}, in presence of exotic charged currents~\cite{Ludl:2016ane} or in relation with KeV neutrino dark matter~\cite{deVega:2011xh,Boyarsky:2018tvu}.

Therefore, we 
investigate the impact of these extensions on the Kurie Plot leading 
 to an information on the effective  electron neutrino mass $m_\beta$, as well as their impact on neutrinoless double beta decay effective mass $m_{ee}$, specially in the case in which KATRIN detects a  discontinuity in the spectrum, meaning  one of the extra sterile fermion mass is below the tritium beta decay threshold $E_0= 18.575$ KeV, and its mixing with the electron neutrino is large enough to be observed  in KATRIN.
This is what is called a {\bf kink} in the beta decay spectrum.%.

 This work is organized as follows:  Section~\ref{sec:experimental}
is devoted to the observables we address that are the tritium beta decay and the neutrino effective Majorana mass in neutrinoless double beta decay, reviewing their experimental (present and future) status.   
A detailed description of the minimal SM extensions with sterile fermion states (i.e. $3+N$, Type-I seesaw and LISS) we
consider, including the parametrization we use for each of them, are presented in Sec.~\ref{sec:model}, while the relevant constraints on the sterile fermions applied in our analysis are summarized  in Appendix~\ref{sec:constraints} (the parametrization for the 3+2 model is summarized in Appendix \ref{sec:para3+2}).
Section~\ref{sec:results} collects our discussion on the 
results (and predictions) obtained for the different models we consider. 
Final conclusions and remarks are given in Sec.~\ref{sec:conclusion}.

\section{Present and future experimental situation}\label{sec:experimental}

Since we are interested by the possible effect that could be observed in tritium beta decay spectrum by KATRIN due to the presence of sterile neutrinos with masses below the threshold of $E_0\sim 18.6$ KeV and also in $0\nu 2 \beta$ effective mass,   we discuss in the following both observables and their associated  present and future experimental sensitivities. 
\subsection{Tritium beta decay experiments}
Analyses of the $\beta$ decay spectrum are the most
model-independent method to directly probe neutrino masses ($m_\beta$) independent of their nature (Dirac or Majorana).
These experiments address the nuclear reaction decay  of 
\begin{equation}
^3 \text{H} \, \to {}^3\text{He} \,+\, e^- + \,
\bar \nu_e \,, 
\end{equation}
the kinematics of which is impacted by the mass of the neutrino leading to a distortion of the electron end-point energy spectrum which depends on the mixings of the  interaction (flavor) eigenstate $\nu_e$  with  the physical eigenstates, $\nu_i$ ($i=1,2,3$)
and on their masses.
 The study of the  electron energy spectrum at the end-point leads to an information on the emitted light neutrino. Given the fact that there is indeed lepton mixing (PMNS), one defines the ``electron effective mass"  as\footnote{Since the mass splittings between the three light mass eigenstates are so small,  the current $\beta$ decay experiments cannot resolve them.} 
 \begin{equation} \label{eq:bdecaymass}
 m_{\beta}\,= \sqrt{\sum_{i=1}^{3} m_i^2 \big| {\bf U}_{ei}\big|^2 }\ , 
 \end{equation}
 where ${\bf U}$ denotes the $3\times 3$ PMNS mixing matrix and where  the sum runs over the three light (active) neutrino physical states with masses $m_i$, $i=1,2,3$. 
 Up to now, the most stringent bounds on $m_\beta$ are those reported
by the Mainz~\cite{Kraus:2004zw} and Troitsk~\cite{Aseev:2011dq}
experiments, 
\begin{align}\label{mainztroisk}
\quad m_\beta \leq 2.3~\text{eV}\,\, (95\% \text{
    C.L.})\,,& \quad  \text{Mainz},\nonumber \\
\quad m_\beta \leq 2.1~\text{eV}\,\, (95\% \text{ C.L.})\, , & \quad  \text{Troitsk}\ ,
\end{align}
while the KATRIN experiment~\cite{Osipowicz:2001sq,Angrik:2005ep} 
aims for a sensitivity of $0.2$~eV ($90\%$ C.L.) after a period of five years (necessary in order to have three years of data-taking) which has  recently  started. 

Moreover, the presence of an additional sterile  fermion with a mixing $U_{e4}$ to the electron neutrino could lead to discontinuities (kinks) in the spectrum. 
This was recently explored by the Troitsk experiment, setting 
limits on $|U_{e4}|$ for a sterile neutrino with a mass of 0.1-2~KeV~\cite{Abdurashitov:2017kka}.
Interestingly,   KATRIN (in its possible future TRISTAN) aims at measuring  the full tritium beta decay spectrum with an unprecedented resolution, allowing them to explore the existence of (at least) one heavy (mostly sterile) neutrino in the mass range of 1-18.5 KeV, with a mixing to the active neutrino $\nu_e$ as\footnote{The sensitivity studies can vary form $10^{-6}$ to $10^{-8}$, depending on the applied technique and on the estimated uncertainties. We will therefore consider $10^{-6}$ as a conservative sensitivity.}  $\left| U_{e4} \right|^2 \geq 10^{-6}$~\cite{Mertens:2015ila,Mertens:2014nha,Mertens:2014osa,Steinbrink:2017ung}, the matrix $U$ being the total lepton mixing matrix. 
Indeed, in the presence of a heavy neutrino with mass $m_4$, the electron energy spectrum would be a superposition of the light neutrino spectrum and the one of the heavy neutrino, both weighted by their corresponding mixing, as follows~\cite{Shrock:1980vy,Mertens:2014osa}
\begin{eqnarray} \label{eq:bdecay}
\frac{d \Gamma}{d E} = \Theta \left( E_0 - E - m_{\beta} \right) \left( 1- \left| U_{e4} \right|^2 \right) \frac{d \Gamma }{d E} \left( m_{\beta}\right) +  \Theta\left( E_0 - E - m_{4} \right) \left| U_{e4} \right|^2 \frac{d \Gamma }{d E} \left( m_4 \right)\ ,
\end{eqnarray}
where $E_0$ is the threshold energy, $E$ is the kinetic electron energy,  $\frac{d \Gamma }{d E}(m)$ is the differential beta spectrum for a neutrino of mass $m$, and where $m_{\beta}$   is the electron effective mass given in Eq.~(\ref{eq:bdecaymass}). 
The Heaviside step functions in Eq.~(\ref{eq:bdecay}) account for energy conservation, since the available energy of the beta decay  has to be large enough to produce the neutrinos.
This discontinuity is expected to be seen in the spectrum, if the mass of the heavy neutrino is below the threshold $E_0$, and if its mixing with the electron neutrino is large enough to be seen in the form of a kink the KATRIN beta spectrum.

\subsection{Neutrinoless double beta decay experimental status}
The observation of a $0\nu 2 \beta$ decay can be interpreted as
being mediated (at tree-level) by massive neutral Majorana fermions, and/or by new interactions and particles fields arising from NP 
models~\cite{Rodejohann:2011mu,Deppisch:2012nb,Bilenky:2014uka,Pas:2015eia,DellOro:2016tmg}.  
In the Standard Model and under the assumption that Majorana neutrinos mediate the $0\nu 2\beta$
decay at tree-level,  the decay amplitude of $0\nu 2 \beta$ is proportional to 
\begin{equation}\label{Gamma0nubb}
\sum G^2_F\, {\bf U}^2_{ei} \, \gamma_\mu \, P_R \, 
\frac{\slash \hspace*{-2.5mm}p + m_i}{p^2 - m_i^2} \, \gamma_\nu \, P_L \,
\simeq \, \sum G^2_F\, {\bf U}^2_{ei} \, \frac{m_i}{p^2} 
\,\gamma_\mu \, P_R \,\gamma_\nu\,,
\end{equation}
where $G_F$ is the Fermi constant, 
$m_i$ the neutrino (physical) mass and $p$ is the neutrino virtual momentum such that $p^2 \simeq - (125 \mbox{ MeV})^2$ (the value corresponds  to  an average of the virtual
momenta in different decaying nuclei). 
Finally the  $0\nu 2 \beta$  decay width  is proportional to the so-called\footnote{The name ``effective electron neutrino Majorana  mass"  is due to the fact that  the first entry of  the squared neutrino mass ($3\times 3$) matrix  in the
interaction basis is given by: 
$m^2_{ee}\, =\,m^2_{\nu_e \nu_e}\, \equiv \, (M^\dagger\,M)_{e e}\, .
$
}
``effective electron neutrino {\bf Majorana} mass'' given by,
\begin{equation}\label{mee}
m_{ee}\,=\,\left|\sum_{i=1}^3 {\bf U}^2_{ei}\,m_i \right|\,.
\end{equation}

Recently, several experiments (among them KamLAND-ZEN~\cite{Gando:2012zm,KamLAND-Zen:2016pfg},
GERDA~\cite{Agostini:2018tnm}, Majorana Demonstrator~\cite{Aalseth:2017btx}, EXO-200~\cite{Auger:2012ar,Albert:2014awa,Albert:2017owj}, CUORE~\cite{Alduino:2017ehq} and CUPID-0~\cite{Azzolini:2018dyb}) have
set strong 
bounds on the  effective mass $m_{ee}$, the most constraining one being provided by the KamLAND-ZEN collaboration~\cite{KamLAND-Zen:2016pfg}
\begin{equation}\label{eq:onu2beta:limit.KZ}
|m_{ee}|\, < \, 0.061-0.165 \text{ eV} \, \, (90\% \, \text{C.L.})\, ,
\end{equation}
where the range is due to the uncertainties on the nuclear matrix elements\footnote{For details concerning the theoretical uncertainties of nuclear matrix elements, see for 
instance~\cite{DellOro:2016tmg,Maneschg:2017mzu}.}.
Regarding future experimental prospects, we present in
Table~\ref{tab:futuresensitivities} the   
sensitivity of ongoing and planned $0\nu 2 \beta$ dedicated experiments.
Note that throughout  our analysis, we take  $|m_{ee}| \simeq 0.01$ eV  as a   representative  value for the 
future sensitivity.

\begin{table}[t!]
\begin{center}
{\begin{tabular}{| l | l | c |}  \hline                       
Experiment & Ref. &  $ |m_{ee} | $ (eV) \\
  \hline
 EXO-200 (4 yr) & \cite{Auger:2012ar,Albert:2014awa} & $0.075 - 0.2$  \\
nEXO (5 yr)  & \cite{Tosi:2014zza,Licciardi:2017oqg}& $0.012 - 0.029$  \\
nEXO (5 yr + 5 yr w/ Ba tagging) &
\cite{Tosi:2014zza} 
& $0.005 - 0.011$   \\
KamLAND-Zen (800~kg)& \cite{Obara:2017ndb}  & $0.025 - 0.080$ \\
KamLAND2-Zen (1000~kg)& \cite{Obara:2017ndb}  & $<0.02 $\\
GERDA  phase II & \cite{Agostini:2017iyd}& $0.09 - 0.29$ \\
MAJORANA DEMONSTRATOR& \cite{Phillips:2011db,Wilkerson:2012ga} & $0.06 - 0.17$ \\
LEGEND  & \cite{Abgrall:2017syy} & $0.011 - 0.023$ \\
CUORE (5 yr) & \cite{Gorla:2012gd,Aguirre:2014lua,Artusa:2014lgv} &
$0.051 - 0.133$ \\ 
CUPID& \cite{Azzolini:2018dyb} &
$0.006 - 0.170$ \\
SNO+  & \cite{Hartnell:2012qd,Lozza:2016ghw} & $0.07 - 0.14$ \\
SuperNEMO & \cite{Barabash:2011aa} & $0.05 - 0.15$ \\
AMoRE-I & \cite{Karki:2018rhc,Jo:2017jod} & $0.12 - 0.2$\\
AMoRE-II & \cite{Jo:2017jod} & $0.017 - 0.03$\\
NEXT & \cite{Granena:2009it,Gomez-Cadenas:2013lta}& $0.03 - 0.1$ \\
 \hline                       
\end{tabular}
}
\caption{Sensitivity of several neutrinoless double beta decay experiments.}
\label{tab:futuresensitivities}
\end{center}
\end{table}

In the situation where the SM is extended by a number $N$ of sterile fermion states,  the additional neutrinos might also contribute to the decay amplitude in which the corresponding effective mass  $m_{ee}$ is corrected~\cite{Benes:2005hn,Blennow:2010th} as follows:
\begin{equation} \label{eq:0vudbdecay}
 m_{ee}\,\simeq \,\sum_{i=1}^{3+N} { U}_{ei}^2 \,p^2
\frac{m_i}{p^2-m_i^2}\ , 
\end{equation}
where $U$ is the $(3+N)\times (3+N)$ lepton mixing matrix and where the sum is done over all the total number of physical neutrino states $n_\nu=3+N$. From the latter expression, we can already notice that  an observation of such a kink in tritium beta decay spectrum (i.e. having one of the extra neutral fermion mass $m_i$ in the $[1-18.5 ]$~KeV range with a mixing to the electron neutrino $\left| U_{ei} \right|^2 \geq 10^{-6}$) would have important consequences on $m_{ee}$, as we will show in this study. \\

\section{Minimal extensions of the SM involving sterile fermions}\label{sec:model}
In order to accommodate neutrino masses and
mixings, the  SM can be extended with new  sterile fermions 
such as RH  neutrinos.
In this work, we consider  the SM with  three light Majorana neutrinos (${\rm {SM_\nu}}$), which is  extended by a number 
$N$ of sterile fermion states that mix with the 3 active neutrinos. 
We first consider that  the neutrino mass eigenvalues and the lepton mixing matrix are independent, meaning that no assumption is made on the neutrino mass generation mechanism.  As we  will see later, we focus on the $3+1~(N=1)$ case and comment on the generalization to the $N\geq2$ cases.
Then, we explore minimalistic but realistic realizations of the Type-I seesaw model with two RH neutrinos. Besides the general case, we will also be interested in the lepton number conserving scenario considering a combination of linear and inverse seesaw model (LISS).

After EW symmetry breaking, the relevant terms in the Lagrangian can be written (in the Feynman-'t Hooft gauge) as,
\begin{eqnarray}
\mathcal{L}\hspace{-0.2cm}&=&\hspace{-0.2cm}
-\frac{g}{\sqrt{2}}U_{\alpha
 i}W_\mu^{-}\overline{\ell_\alpha}\gamma^{\mu}P_L\nu_i
-\frac{g}{\sqrt{2}}U_{\alpha
i}H^{-}\overline{\ell_\alpha}\left(\frac{m_\alpha}{m_W}P_L-\frac{m_i}{m_W}P_R\right)\nu_i 
+\mathrm{H.c.}\nonumber\\
&&\hspace{-0.2cm}
-\frac{g}{2\cos\theta_W}U_{\alpha i}^*U_{\alpha
j}Z_{\mu}\overline{\nu_i}P_L\nu_j
-\frac{ig}{2}U_{\alpha i}^*U_{\alpha j}A^0\overline{\nu_i}\left(\frac{m_j}{m_W}P_R\right)\nu_j
+\mathrm{H.c.}\nonumber\\
&&\hspace{-0.2cm}
-\frac{g}{2}U_{\alpha i}^*U_{\alpha
j}h\overline{\nu_i}\left(\frac{m_j}{m_W}P_R\right)\nu_j
+\mathrm{H.c.}\ , 
\label{eq:lag}
\end{eqnarray}
where $g$ is the $SU(2)_L$ gauge coupling, $U_{\alpha i}$ are the 
 lepton mixing matrix components, 
$m_i$  are the mass eigenvalues of the neutrinos and  $m_\alpha$ are the charged lepton masses. 
The indices $\alpha$ and $i$ run as  $\alpha=e,\mu,\tau$ and $i=1,\dots,3+N$. Further details can be found in for example
Refs.~\cite{Ilakovac:1994kj, Alonso:2012ji}. 

We proceed first by presenting the 3+$N$ models and the parametrization we used for the minimal cases of $N=1$ and $N=2$. We then detail the low-scale minimal seesaws we consider in this work, i) the Type-I seesaw with two right-handed neutrinos without any hypothesis on the degeneracy of their mass, meaning no lepton number conservation symmetry is imposed, ii) still with two fermionic singlets, we take the limit of the latter mechanism with a small lepton number violation, i.e., a combination of the linear and the inverse seesaw mechanisms. For all the scenarios we detail the corresponding  parametrization we adopt in the numerical study.

\subsection{Effective $\pmb{3+ N}$ models}
\label{sec:EFF}
Since the generic idea of having impact on our observables
  applies to any model
where the active neutrinos have sizable mixings with the additional sterile fermions,  
we can use an \textit{effective} model with 3 light active neutrinos plus N extra sterile neutrinos.

In this framework 
the leptonic charged current is modified as
\begin{equation}\label{eq:cc-lag}
- \mathcal{L}_\text{cc} = \frac{g}{\sqrt{2}} U_{ji} 
\bar{\ell}_j \gamma^\mu P_L \nu_i  W_\mu^- + \, \text{H.c.}\,,
\end{equation}
where 
$i$ denotes the physical neutrino states, from 1 to $n_\nu=3+N$, 
and $j = 1, \dots, 3$ the flavor of the charged leptons. 
In the case of three neutrino generations,  $U$ corresponds
to  the (unitary) PMNS matrix. 
For $n_\nu\geq4$, the 
mixing between the left-handed leptons, which we will subsequently 
denote by $\tilde U_\text{PMNS}$, 
corresponds to a $3 \times 3$ block of $U$. 
One can parametrize the 
$\tilde U_\text{PMNS}$ mixing matrix as
\begin{equation}\label{eq:U:eta:PMNS2}
\tilde U_\text{PMNS} \, = \,(\mathbb{\1} - \zeta)\, 
U_\text{PMNS}\,,
\end{equation}
where the  matrix $\zeta$ encodes the deviation of $\tilde
U_\text{PMNS}$ from unitarity, due to the presence of sterile fermions.
Given the modification of the charged current in Eq.~(\ref{eq:cc-lag}), many observables will be sensitive to the 
active-sterile mixings, and their current experimental values (or
bounds) will thus constrain such an extension. These constraints arise from lepton flavor violating (and universality violating) observables, bounds from  laboratory
and collider searches, among others.
Certain
sterile mass regimes and active-sterile mixing angles are also strongly constrained by cosmological observations. All the relevant constraints for the mass regimes we consider in this study  are discussed in Appendix \ref{sec:constraints}.

Note that in the $3+N$ model, the mixing matrix $U$ includes $(3+N)(2+N)/2$ rotation angles,
$(2+N)(1+N)/2$ Dirac phases and $2+N$ Majorana phases. All these constitute the physical parameters of 
the model in addition to the masses of the sterile states, $m_i$, $i=1,\dots, N$.

\subsubsection{Mixing matrix  $\pmb{3+N}$: parametrization}
We have conducted the study for the most minimal cases $N=1$ and $N=2$. In the $3+1$ model,  the introduction of the extra sterile state reflects into three new mixing angles ($ \theta_{14}, \theta_{24}, \theta_{34}$) (active-sterile mixing angles), two extra Dirac CP violating phases ($\delta_{41},\delta_{43}$) and an extra Majorana phase ($\phi_{41}$). The 4$\times$4 lepton mixing matrix is now given by the product of 6 rotations times the Majorana phases\footnote{We recall that since we are interested in the impact of sterile fermions on neutrinoless double beta decay effective mass, we assume in the whole study that  neutrinos are of Majorana nature.}:
\begin{eqnarray} \label{eq:3+1rot}
%\beq
U &=& R_{34}(\theta_{34},\delta_{43}) \cdot R_{24}(\theta_{24}) \cdot R_{14}(\theta_{14},\delta_{41}) 
\cdot R_{23} \cdot R_{13} \cdot R_{12}  \cdot \rm diag(1,e^{i \phi_{21}},e^{i \phi_{31}},e^{i \phi_{41}}) \nonumber \\
&=& R_{34}(\theta_{34},\delta_{43}) \cdot R_{24}(\theta_{24}) \cdot R_{14}(\theta_{14},\delta_{41}) 
\cdot U_{\rm PMNS}^{4\times4} \cdot \rm diag(1,e^{i \phi_{21}},e^{i \phi_{31}},e^{i \phi_{41}})\ ,
\end{eqnarray} 
where $U_{\rm PMNS}^{4\times4}$ is the $4\times4$ matrix formed by the $3\times3$  PMNS matrix,  which is extended with a trivial fourth line and a fourth  column,  and where 
the rotation matrices $R_{34},R_{24},R_{14}$ are defined as:
\begin{eqnarray} \label{eq:R}
R_{14}\ &=& \left( 
\begin{array}{cccc}
\rm cos \theta_{14} & 0 & 0 & \rm sin \theta_{14} \cdot e^{-i \delta_{41}} \\
0 & 1 & 0 & 0 \\
0 & 0 & 1 & 0 \\
- \rm sin \theta_{14} \cdot e^{i \delta_{41}} & 0 & 0 &\rm cos \theta_{14} \\
\end{array}
\right), R_{24}\ =\  \left( 
\begin{array}{cccc}
1 & 0 & 0 & 0 \\
0 & \rm cos \theta_{24}  & 0 & \rm sin \theta_{24}\\ 
0 & 0 & 1 & 0 \\
0 & - \rm sin \theta_{24}& 0 &  \rm cos \theta_{24} %
\end{array}%
\right)\ , \nonumber \\
R_{34}\ &=&\ \left( 
\begin{array}{cccc}
1 & 0 & 0 & 0 \\
0 & 1 & 0 & 0 \\
 0 & 0 & \rm cos \theta_{34} &\rm sin \theta_{34} \cdot e^{-i \delta_{43}}\\ 
 0 & 0 & -\rm sin \theta_{34} \cdot e^{i \delta_{43}}& \rm cos \theta_{34} 
\end{array}%
\right) \ .
\end{eqnarray} 
The parametrization for the lepton mixing matrix for the $3+2$ model ($N=2$) is shown in Appendix \ref{sec:para3+2}.

\subsection{Minimal seesaw mechanisms with two sterile fermions}
\subsubsection{Type-I seesaw with two right-handed neutrinos and parametrization}
In order to comply with neutrino data,  the most minimal realization of the Type-I seesaw mechanism requires only  two right-handed neutrinos, meaning that the lightest active neutrino is massless. 
The Lagrangian of the Type-I seesaw reads
\bee\label{eq:lag_ssI}
\mathcal{L} &=& \mathcal{L}_\text{SM} + i \overline{N_I} \slashed{\partial} N_I - \left( Y_{\alpha I} \overline{\ell_\alpha} \tilde{\phi} N_I + \frac{M_{IJ}}{2} \overline{N_I^c} N_J + \text{H.c.}\right),
\eee
where $\ell_\alpha$ are the SM lepton doublets, $\phi$ is the Higgs doublet  and $\tilde\phi = i\sigma_2\phi^*$, $N_I$ denotes the new fermionic fields that are singlet under the SM gauge group, $Y_{\alpha I}$ are dimensionless Yukawa couplings and $M$ is a  $2\times 2$ matrix of Majorana mass terms for the $N_I$ fermions. 
Without loss of generality, we will assume $M$ to be diagonal.

After the EW symmetry breaking the Higgs field acquires a non-vanishing vacuum expectation value (VEV) $ \<\phi\>=v$ ($174$ GeV), and the full neutrino mass matrix in the EW basis can be written as follows 
\be\label{eq:mTypeI}
M_{\rm Type-I}=\left(\begin{array}{cc} 0 & m_D \\ m_D^T & M \end{array}\right)\,,
\ee
where  $m_D$ denotes the   $3\times 2$ Dirac mass  matrix, $m_{D_{\alpha I}}= v\ Y_{\alpha I}$.  The Lagrangian~(\ref{eq:lag_ssI}) accounts for a non-vanishing (active) neutrino mass matrix $m_\nu$ which, after the block diagonalisation of the matrix $M_{\rm Type-I}$ and under the assumption $v |Y_{\alpha I}| \ll |M_{IJ}|$ (seesaw limit), is given by 
\bee\label{eq:numSS}
m_\nu=m_{\text{light}} \simeq - {v^2}{} Y {M}^{-1} Y^T\ . \eee
For our numerical study, we adopt the following parametrization for the above defined Dirac mass (details can be found in Refs.~\cite{Casas:2001sr,Donini:2011jh,Antusch:2009pm}): 
\begin{equation}\label{paramtypeI}
m_D^T = i \sqrt{M} R\  \textbf{U}^{\dagger}\ ,\end{equation}
where $\textbf{U}$ is the PMNS matrix. Depending on the ordering in the light neutrino spectrum (inverted or normal ordering that we label IO or NO, respectively) and given the fact that in this minimal scheme with only two RH neutrinos, one active neutrino is massless, $m_{\text{lightest}}=0$, the matrix $R$ is such that $R^T R$ is the diagonal light neutrino mass matrix for each ordering. 
This respectively corresponds to
\begin{eqnarray}
\text{NO}:\ \  &&R=R_{NO} =  
\left(
\begin{matrix}
0 & \sqrt{m_2} \cos(a+ib) & \sqrt{m_3} \sin(a+ib) \\
0 & \mp \sqrt{m_2} \sin(a+ib) & \pm \sqrt{m_3} \cos(a+ib)
\end{matrix}
\right)\ ,
\label{R parametrization NO}\\
\text{IO}:\ \  &&R=R_{IO} = 
\left(
\begin{matrix}
\sqrt{m_1} \cos(a+ib) & \sqrt{m_2} \sin(a+ib) & 0\\
\mp \sqrt{m_1} \sin(a+ib) & \pm \sqrt{m_2} \cos(a+ib) & 0
\end{matrix}
\right)\ ,
\label{R parametrization IO}
\end{eqnarray}
where $a,b \in \mathbb{R}$ and where $m_1$, $m_2$ and $m_3$ are the light neutrino masses satisfying the solar and atmospheric mass squared splittings, $\Delta m^2_{\text{sol}}$ and $\Delta m^2_{\text{atm}}$.

\subsubsection{Approximate lepton number symmetry: Linear and Inverse seesaw with 2 sterile fermions}
Among the several variation of the low-scale seesaws,   the Inverse or the Linear seesaw mechanisms do offer the possibility of having the heavy neutrinos in pairs forming pseudo-Dirac states. 
These mechanisms are based on approximate lepton number symmetry, in which the smallness of the neutrino masses is related to the smallness of LNV parameters, which are natural in the sense of 't Hooft~\cite{tHooft:1979rat}, since the Lagrangian acquires a new symmetry when they are set to zero, making  therefore neutrino masses stable against radiative corrections. In addition, the small mass splitting between the two states of each pair (i.e. strong degeneracy in mass) is proportional to the source of LNV. 

The minimal setup in this mechanisms is to extend the SM with a pair of sterile fermions, $N_{1,2}$, with opposite lepton number, $L=\pm 1$. In the case with only one active generation, the lepton number conserving part of the neutrino mass matrix reads, in the basis $(\nu_L, {N_1}^c, {N_2}^c)$,
\bee\label{eq:M0_toy}
{M}_0 &=& \left(\begin{array}{ccc} 0 & y v & 0\\
y v & 0 & \Lambda \\
0 & \Lambda & 0
\end{array}\right),
\eee  
where $y$ is a dimensionless Yukawa coupling,  $\Lambda$  a dimension-full parameter, and $v$ the Higgs VEV. 
The lepton number conserving mass spectrum resulting from the diagonalisation of this mass matrix is composed by a massless state $m_\nu \equiv M_1= 0$ and two degenerate Majorana massive states, $M_2 = M_3 = \sqrt{\Lambda^2 + v^2 y^2}$.
In order to account for massive light (Majorana)  neutrinos, one has to consider a correction to the latter mass matrix  by adding small LNV entries. Forbidding  a non-zero element in the $(1,1)$ entry, which would correspond to a Majorana mass term for left-handed neutrinos and requires a non-minimal extension of the SM (instance a Type-II seesaw), there are two possibilities\footnote{Actually there is a third one corresponding to having a non-vanishing $(2,2)$ entry leading to an Extended (radiative) seesaw generating neutrino masses  only at higher loop level that can gain importance only  in the case of a  large lepton number violation. Since we are interested in a possible double-kink in KATRIN, which would be associated to very small LNV in this context, we keep  the $(2,2)$ entry in Eq. (\ref{eq:M0_toy}) to zero.}  resulting to the patterns of inverse and linear seesaws
\begin{equation}
\begin{array}{lccr}
\Delta M_{\rm ISS}=\left(\begin{array}{ccc}
 0 & 0 & 0\\
0 & 0 & 0\\
0 & 0 & \xi \Lambda
\end{array}\right),& \hspace{-0.03\textwidth} 
&
\Delta M_{\rm LSS}=\left( \begin{array}{ccc} 0 & 0 & \epsilon y v\\
0 & 0 & 0\\
\epsilon\, y v & 0 & 0 
\end{array}\right),& \hspace{-0.03\textwidth} 
\end{array}
\label{eq_pert}
\end{equation}
where $\xi$ and $\epsilon$ are small $(<1)$ dimensionless LNV parameters. After diagonalisation of $M_0 +\Delta M$, the mostly active neutrino mass $m_\nu$ for each mechanism, at  leading order in $\xi$ and $\epsilon$ are 
\begin{align}\label{eq:mnu}
\text{ISS}:  m_\nu= \xi y^2 \frac{v^2}{\Lambda}\ ,   \quad
 \text{LSS}: m_\nu= 2 \epsilon y^2 \frac{v^2}{\Lambda}\ .
\end{align}
In this work, we will assume the existence of two sterile neutrinos and consider both sources of LNV small corrections, naming the model  ``LISS" : LSS+ISS.

In the realistic case of 3 active generations,  the mass matrix for the LISS model is given by~\cite{Gavela:2009cd}
\begin{equation}\label{eq:MLISS}
M_{LISS} =  \left( \begin{array}{ccc}
 \mathbf{0} & \textbf{Y} v &\epsilon \textbf{Y}' v\\
 \textbf{Y}^T v & 0 & \Lambda \\
 \epsilon{\textbf{Y}}'^T v & \Lambda & \xi \Lambda 
 \end{array}
\right),
\end{equation}
where $\textbf{Y}$ is now a 3-dimensional vector providing the Dirac mass for the active neutrinos $v \textbf{Y}= m_D$. 
Notice that the ordering of the second and third column/row  of Eq.~(\ref{eq:MLISS}) is due to the assignment $L=+1$ and $-1$, for $N_1$ and $N_2$, respectively.

\subsection{LISS parametrization}
To ease our analysis and parameter counting,  we set $\mu\equiv \xi \Lambda$, $\varepsilon\equiv \epsilon \textbf{Y}' v $ in  $M_{\rm LISS}$ defined in Eq.~(\ref{eq:MLISS}): 
\begin{equation}
M_{LISS} =
\left( 
\begin{matrix}
0 & m_D & \varepsilon \\
m_D^T & 0 & \Lambda \\
\varepsilon^T & \Lambda & \mu
\end{matrix}
\right)\ .
\label{eq:MLISSBIS}
\end{equation}  
In the seesaw limit, where $\left| m_D \right|, \left| \varepsilon \right|, |\mu| \ll \Lambda$, the block diagonalisation of the latter leads  to 
\begin{eqnarray*}
U_B^T M_{\rm LISS} U_B = 
\left(
\begin{matrix}
m_{\text{light}}^{3\times 3} & \begin{matrix}
0_{3\times1 } & 0_{3\times1 }
\end{matrix} \\
\begin{matrix} 0_{1 \times 3} \\ 0_{1\times 3 } \end{matrix} & M_{\text{heavy}}^{2\times 2}
\end{matrix}
\right)\ ,
\label{LISS block diagonalized}
\end{eqnarray*}
where $U_B$ is a unitary matrix,  and where $m_{\text{light}}$  and $M_{\text{heavy}}$ are given as follows,
\begin{eqnarray}\label{LISS-mlight}
&&m_\nu\equiv m_{\text{light}} \simeq \frac{1}{\Lambda} \left( \mu \frac{m_D^T m_D}{\Lambda} - \left( m_D^T \varepsilon + \varepsilon^T m_D \right) \right)
\ ,\\
&&M_{\text{heavy}} \simeq 
\left(
\begin{matrix}
0 & \Lambda  \\
\Lambda & \mu
\end{matrix}
\right)\ .
\label{Mheavy}
\end{eqnarray}
Notice that we take $M_{\text{heavy}}$ at zeroth order since the degeneracy is already broken with the mass term $\mu$. Identifying~\cite{Gavela:2009cd} 
$\varepsilon' = \varepsilon - \frac{\mu}{2 \Lambda}$ in Eq.~(\ref{LISS-mlight}), the (mostly active) light neutrino mass $m_{\nu}$ can be rewritten as:
\begin{equation}
m_{\nu} = - \frac{m^T_D \varepsilon' + \varepsilon'^T m_D}{\Lambda}\ .
\label{LISS mlight rewriten}
\end{equation}

\noindent Imposing that $m_\nu$ complies with neutrino data (PMNS mixings and solar and atmospheric mass squared differences), $m_{\nu} =   \textbf{U}^* \text{Diag}\{m_1, m_2, m_3 \}   \textbf{U}^{\dagger}$ (where $\textbf{U}\equiv U_{\text{PMNS}}$
), we obtain for the two different orderings of the light neutrino mass spectrum:  
\hspace*{-1cm}\begin{eqnarray}
\hskip -0.5cm & m_{Dj}^\text{NO} = \eta \frac{\sqrt{\Lambda}}{\sqrt{2}} \left( \sqrt{m_3} \textbf{U}^*_{j3}+ i \ \sqrt{m_2} \textbf{U}_{j2}^* \right) \!\text{; }
& \varepsilon_{j}^\text{NO} = \frac{1}{\eta} \frac{\sqrt{\Lambda}}{\sqrt{2}} \left( \sqrt{m_3} \textbf{U}^*_{j3} - i \ \sqrt{m_2} \textbf{U}^*_{j2} \right) + \frac{\mu}{2 \Lambda} \ ,\nonumber \\
\hskip -0.5cm & m_{Dj}^\text{IO} = \eta \frac{\sqrt{\Lambda}}{\sqrt{2}} \left( \sqrt{m_2} \textbf{U}^*_{j2} + i \ \sqrt{m_1} \textbf{U}^*_{j1} \right)  \!\text{; }
& \varepsilon_{j}^\text{IO} = \frac{1}{\eta} \frac{\sqrt{\Lambda}}{\sqrt{2}} \left( \sqrt{m_2} \textbf{U}^*_{j2} - i \ \sqrt{m_1} \textbf{U}^*_{j1} \right)  + \frac{\mu}{2 \Lambda}\ , 
\label{LISS m_D solution}
\end{eqnarray}
where $\eta$  is a real parameter such that $\left| m_D \right|, \left| \varepsilon \right| \ll \Lambda$.

Finally, the heavy mass matrix Eq.~\eqref{Mheavy} eigenstates are given by
\begin{equation}
m_{4,5} \simeq \Lambda \pm \frac{1}{2} |\mu| \ . 
\end{equation}

 One could hope to have the lepton number parameter $\mu$, which obviously breaks the degeneracy in the mass of the two mostly sterile states, and the mixing between the two heavy states such that KATRIN would see a double-kink, provided  their mixings to the electron neutrino  both lie within its sensitivity, as we will discuss in Section~\ref{sec:LISSresults}. 

%%%%%%%%%%%%%%%%%%%%%%%%%%%%%%%%%%%%%%
%%%%%%%%%%%%%%%%%%%%%%%%%%%%%%%%%%%%%%

\section{Numerical results and discussion}\label{sec:results}
We work under the hypothesis that KATRIN will see a kink in the $\beta$ spectrum. This signal would imply the existence of at least\footnote{In case where the spectrum reveals more than one kink, i.e., more than one sterile neutrino, we assume that the KATRIN sensitivity on $|U_{ei} |$ is the same for $i=4, 5, ...$ Nevertheless, a more dedicated study under the assumption of more than one kink is needed.}  a fourth neutrino - under the hypothesis that the SM should be most minimally extended -  with a mass and mixing to the electron of~\cite{Mertens:2014nha,Mertens:2014osa,Mertens:2015ila,Steinbrink:2017ung}
\begin{equation}\label{KATRINsensitivity}
m_4 \in [1~{\rm KeV}, 18.5~{\rm KeV}]\,, \quad \big|U_{e4}\big|^2 > 10^{-6}\,.
\end{equation}
In the case where the extra neutral leptons are of Majorana nature, one can explore their impact on $0\nu 2 \beta$.
Our aim is thus to study if the interplay between the two observables, the electron energy spectrum in  $\beta$ decay in Eq.~(\ref{eq:bdecay}) and  the Majorana effective mass defined in Eq.~(\ref{eq:0vudbdecay}),  can help discriminating  between motivated low-scale seesaw realizations involving at least one KeV sterile neutrino. 

In our study we consider most minimal extensions of the SM involving at least one sterile neutrino with mass $m_4$ and mixing $U_{e4}$ within the future KATRIN sensitivity.
When several sterile fermions are present, we will assume that $\nu_4$ is the lightest one\footnote{We have also explored scenarios  with an eV and a KeV sterile neutrino and found viable solutions within the minimal seesaw models. Albeit this situation could be interesting for neutrino oscillations anomalies, our discussion on $\beta$ and $0\nu 2 \beta$ decays would not change.}.
 We first consider the {\it ad-hoc} scenario of $3+N$, where the SM is extended by $N$ sterile fermions without any assumption on the neutrino mass mechanism.
 Then, we consider the Type-I seesaw with 2 $\nu_R$ and the LISS scenario (see Section \ref{sec:model}) as explicit examples to show how the interplay between $\beta$ and $0\nu 2 \beta$ decays would affect neutrino mass generation models. 

In the case of the $3+1$ toy model, $m_{ee}$ is given by
\begin{equation}\label{mee3p1}
m_{ee}^{(3+1)}= \sum_{i=1}^4 U_{ei}^2\, p^2 \frac{m_i}{p^2-m_i^2} \simeq \sum_{i=1}^4 U_{ei}^2\, m_i\equiv m_{ee}^{(\rm {SM_\nu})} + m_4 U_{e4}^2\, ,
\end{equation}
$m_4$ being below the nuclear scale $p^2\sim -(125~{\rm MeV})^2$, and more specifically within KATRIN range in Eq.~(\ref{KATRINsensitivity}), and $m_{ee}^{(\rm {SM_\nu})}$ being the effective mass in the $\rm {SM_\nu}$ involving massive Majorana neutrinos according to oscillation data, as defined in Eq.~(\ref{mee}).
In the case of a second sterile neutrino in the $3+2$  model (still no hypothesis of the neutrino mass generation mechanism), the Majorana effective mass can be written as follows:
\begin{equation}\label{mee3p2}
m_{ee} = \sum_{i=1}^5 U_{ei}^2\, p^2 \dfrac{m_i}{p^2-m_i^2}\simeq 
 m_{ee}^{(3+1)}+ U_{e5}^2\, m_5 \frac{p^2}{p^2-m_5^2}\,.
\end{equation}
Notice that, contrary to $m_4$, we are not imposing $m_5$ to be within the KATRIN regime, but we let it  as a free parameter.
Depending on the ranges for $m_5$ and $U_{e5}$, one could have sizable contributions to the neutrinoless  double beta decay effective mass, or even have a cancellation, depending also on the light neutrino spectrum ordering. 
However, when the extra masses and couplings are interdependent due to the embedding of a seesaw, one could have a completely different picture~\cite{Blennow:2010th}. 
For instance, in the case of the Type-I seesaw with 2 $\nu_R$, the neutrino mass diagonalisation requires the condition 
\begin{equation}\label{eq:seesaw.condition}
\sum_{i=1}^5 U_{ei}^2\, m_i =0 \,,
\end{equation}
implying that,
\begin{equation}\label{Usqm5}
U_{e5}^2\, m_5 = -\sum_{i=1}^4 U_{ei}^2\, m_i = - m_{ee}^{(3+1)}\,.
\end{equation}
Using this equation in Eq.~(\ref{mee3p2}), the full effective mass can therefore be written as
\begin{equation}\label{eq:meeseesaw}
m_{ee} \simeq 
 m_{ee}^{(3+1)}\times  \bigg[1-\frac{p^2}{p^2-m_5^2}\,\bigg]\,.
\end{equation}
Interestingly, the last expression in  Eq.~(\ref{eq:meeseesaw}) exhibit two limits, it vanishes if $m_5^2 \ll |p^2|$ and goes to the (3+1) case in the $m_5^2\gg |p^2|$ decoupling limit, as we will address in our numerical results.

It is worth mentioning that  for any seesaw model involving $N$ sterile states, when all their masses are below the threshold of any lepton number violating processes (in for instance mesons and tau lepton LNV decays leading to same or different flavor and same electric charge leptons $ee,\ e\mu, \ \mu\mu,\ e\tau,$ ... ), one can generalize the discussion above on the Majorana electron effective mass $m_{ee}$ to a $3\times 3$ Majorana flavor effective mass 
 \begin{equation}\label{mflavor3x3}
\mathcal M_{\alpha\beta}\equiv\sum_{i=1}^{3+N} U_{\alpha i }^* U_{\beta i}^* m_i \,,
\end{equation}
$\mathcal M$ being the mass matrix in the flavor basis (whose $(1,1)$ entry is $m_{ee}$)~\cite{Abada:2017jjx}. 
 Eq.~(\ref{mflavor3x3}) is related to the $\nu_L$-$\nu_L$ entry in the neutrino mass matrix, which is zero in the Type-I seesaw model, see Eq.~(\ref{eq:mTypeI}), implying 
 \begin{equation}
 \mathcal M_{\alpha\beta}= 0\,,
 \end{equation}
 for Type-I seesaw models with all  sterile neutrino masses  below the energy threshold of the associated LNV process. 
 This is a generalization of the GIM-like cancellation for $m_{ee}$ discussed in~\cite{Blennow:2010th}.

Regarding the numerical analysis for the different seesaw models we consider, we use their corresponding parametrization (detailed in Section \ref{sec:model}) and perform a ``random" scan on all the parameters including  the CP violating phases. We impose that the outcome of the diagonalisation of the mass matrices for the light neutrino parameters, masses and mixings, must lie within $5\%$ from the current best fit values that we take from the global analysis of~\cite{Esteban:2016qun} for the normal and inverted  ordering, whereas we apply,  when relevant, all the constraints detailed in Appendix~\ref{sec:constraints} on  the heavy sector parameter space.  It is worth mentioning that given the mass regimes for the heavy neutrinos we consider, the most constraining bounds are from direct search constraints. 
Note that since we are interested in probing the KeV sterile neutrino hypothesis by laboratory searches independently of cosmology, the cosmology constraints are not applied in our numerical analysis. Nevertheless, we discuss in Appendix~\ref{sec:constraints} the most relevant cosmological constraints and possible mechanisms to avoid them.

\subsection{Results for the $\pmb{3+N}$ model }
We consider first the case where only one sterile neutrino is added to the SM field content and assume it to be within KeV mass range. 
In this case, besides the possibility of having a potential signal (kink) in the beta energy spectrum, the sterile neutrino gives a further contribution to  the $0\nu 2 \beta$ effective mass according to Eq.~(\ref{mee3p1}).

\begin{figure}[t!]
\begin{center}
\begin{tabular}{cc}
 \includegraphics[width=0.49\textwidth]{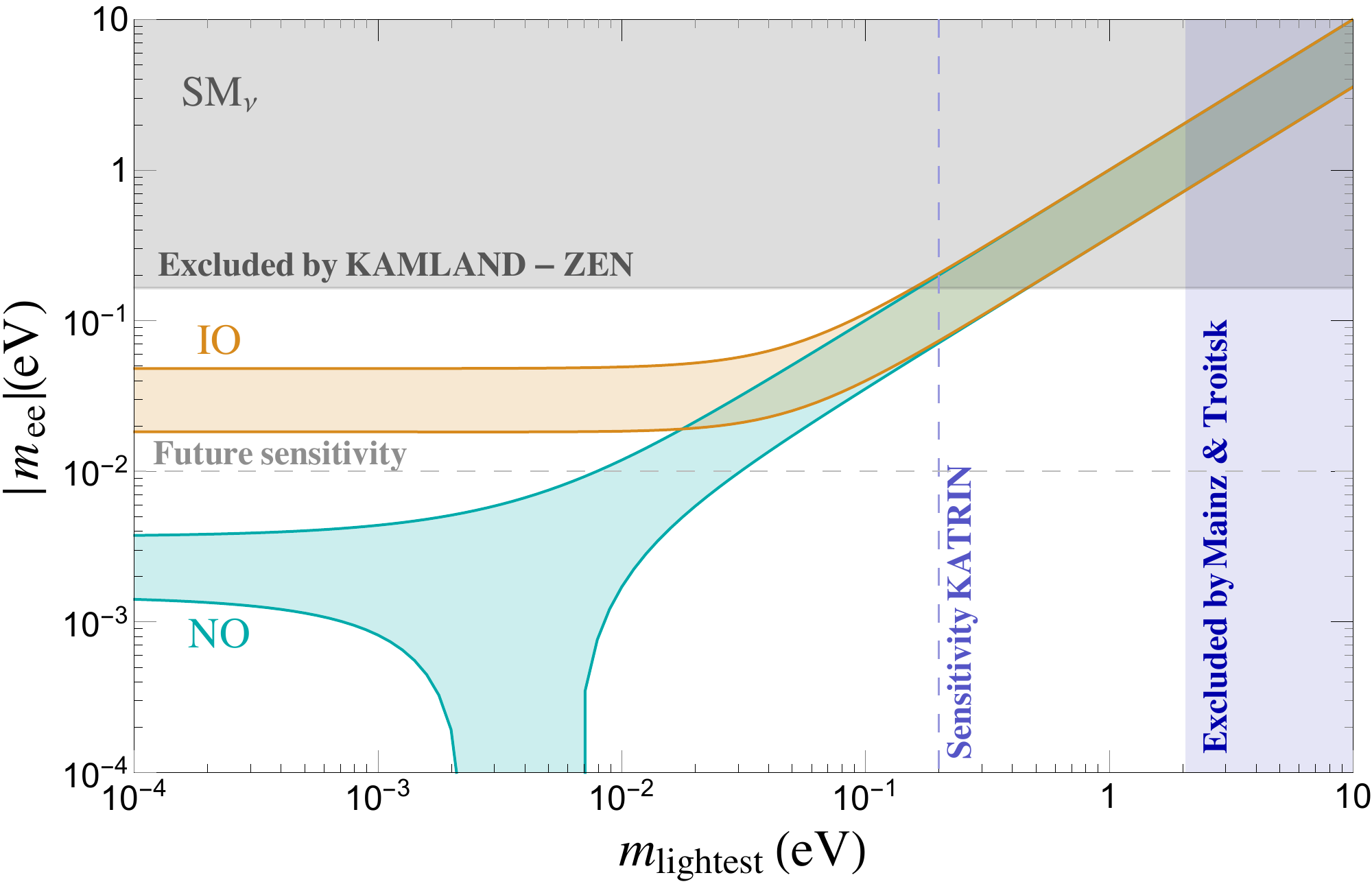}  &  \includegraphics[width=0.49\textwidth]{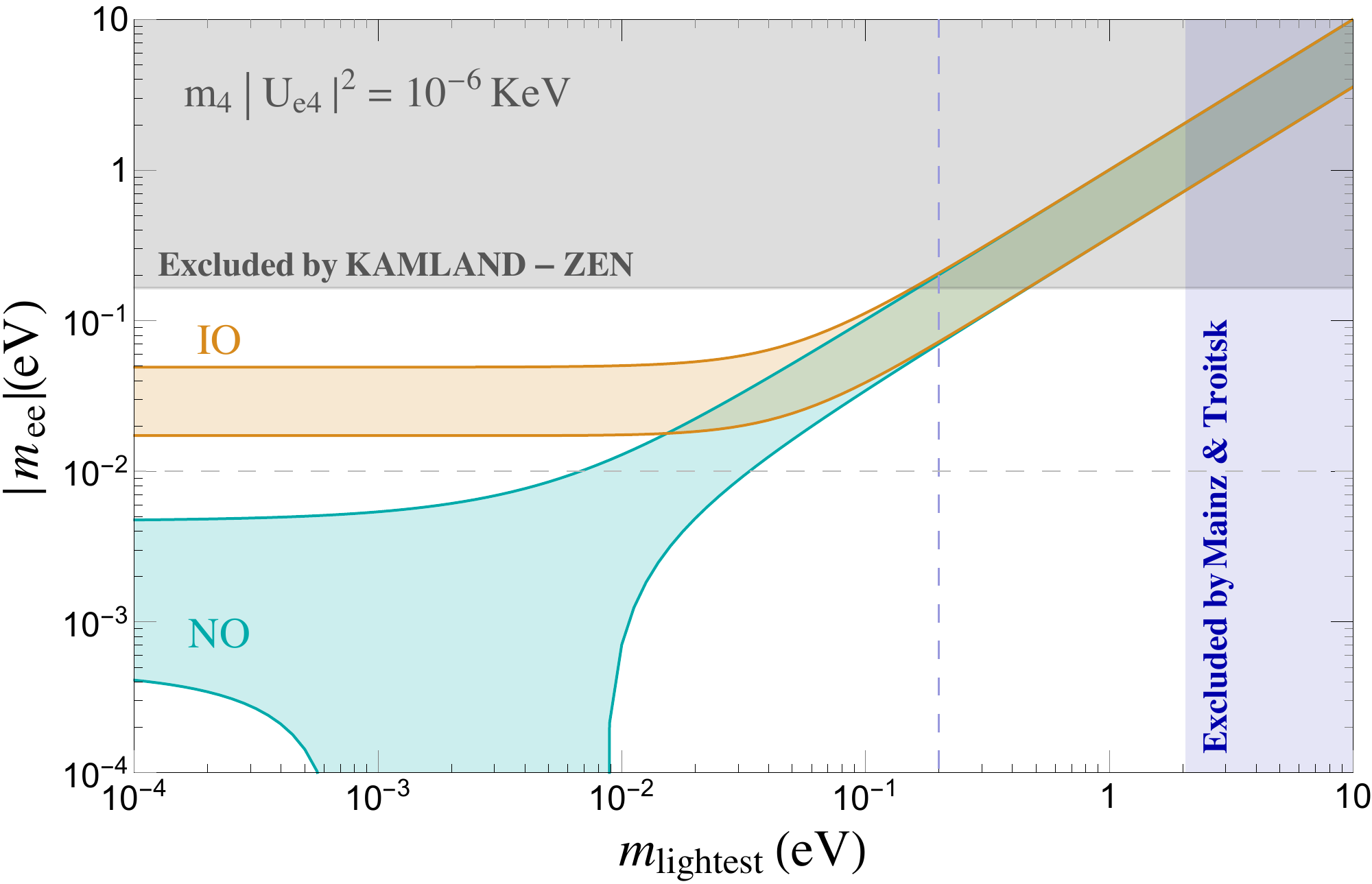}
 \\(a)&(b)\\
 \includegraphics[width=0.49\textwidth]{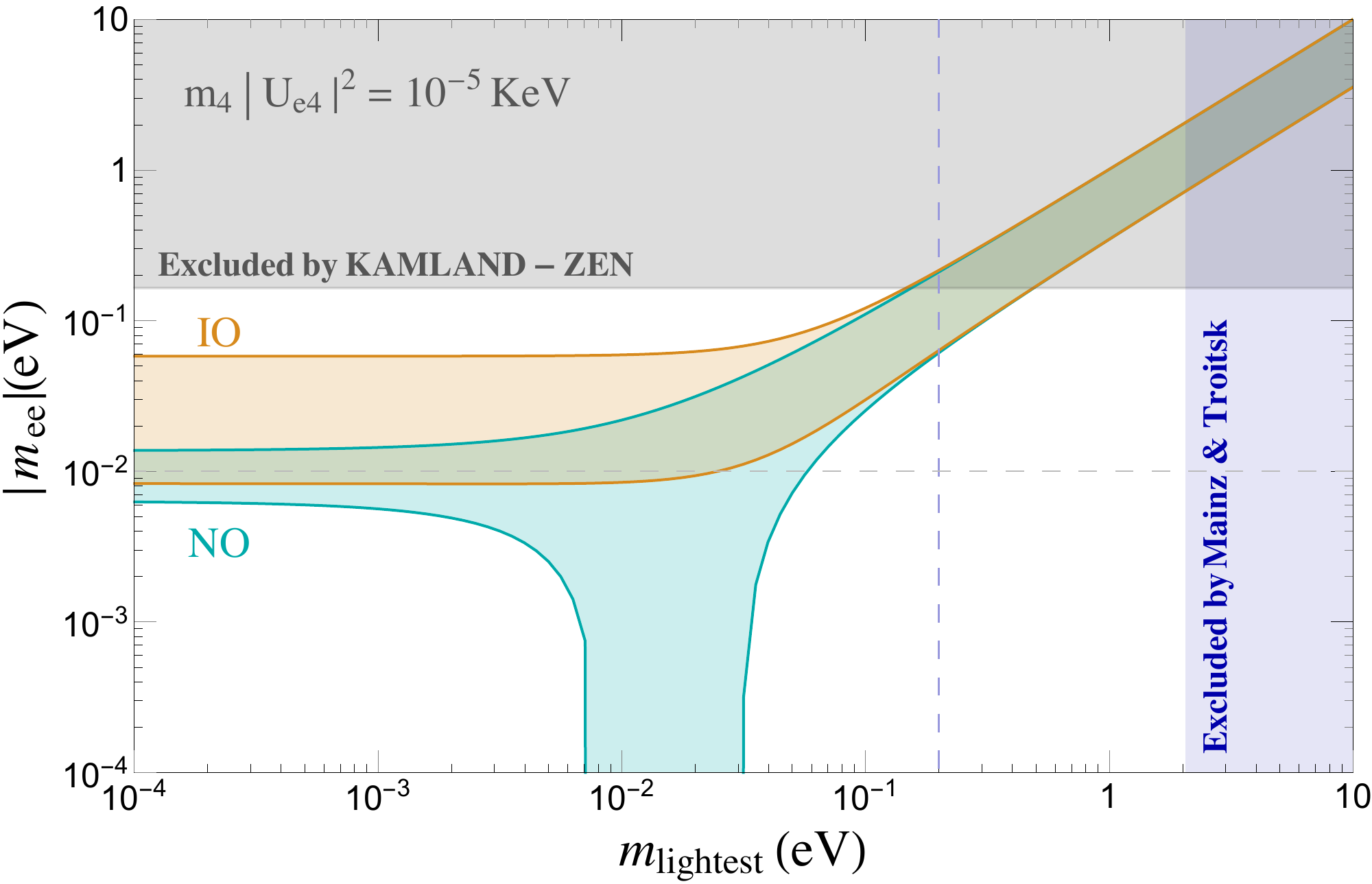}
 & 
  \includegraphics[width=0.49\textwidth]{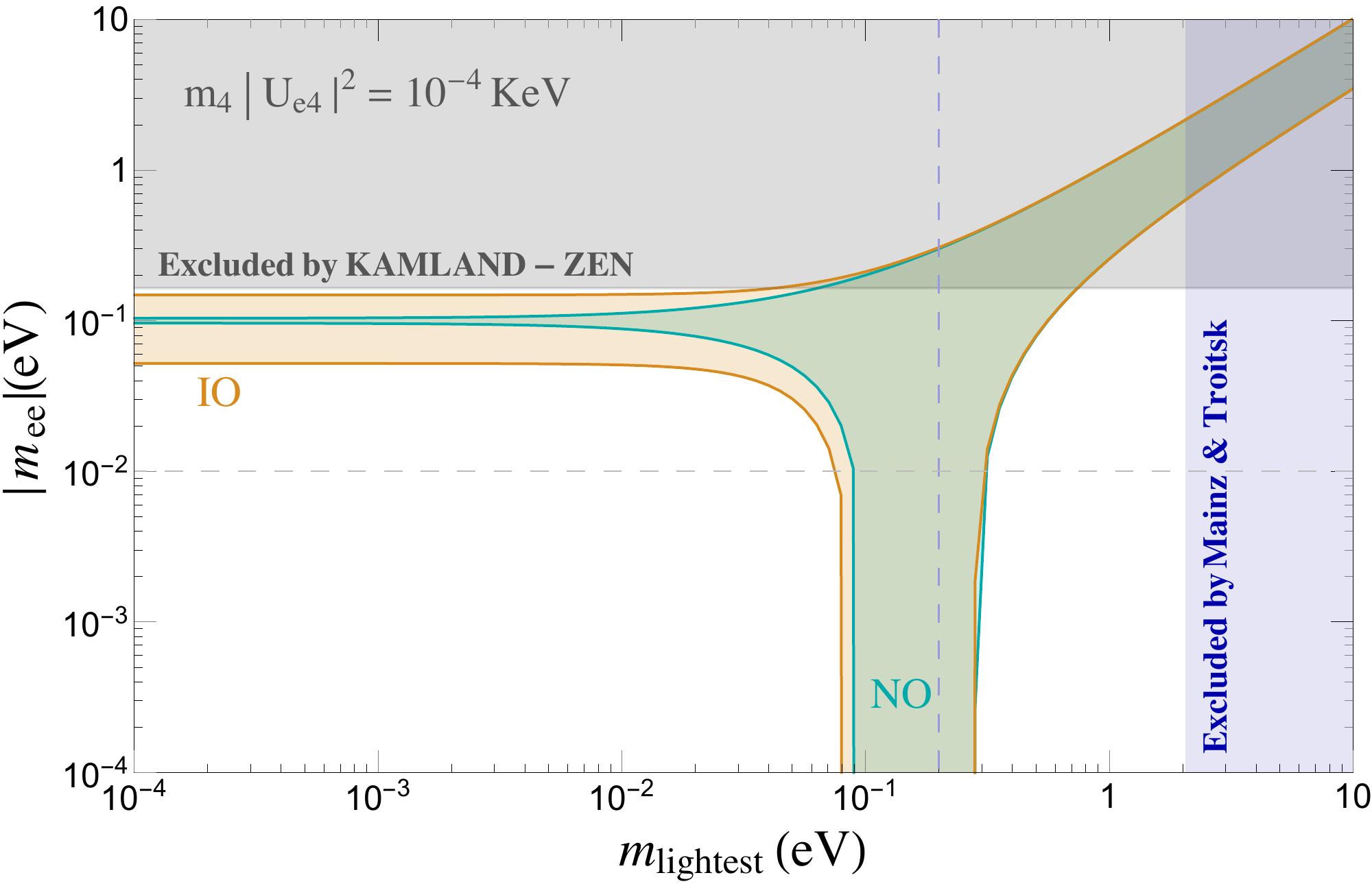} \\(c)&(d)\end{tabular}
\end{center}
 \caption{Effective Majorana  mass $|m_{ee}|$ as a function of the lightest active neutrino mass. The present situation of the SM with the three active neutrinos (${\rm SM}_\nu$) is presented in (a). The other three figures correspond to the situation of the 3+1 model for three representative cases for $m_4|U_{e4}|^2$: $10^{-6}$ (b), $10^{-5}$ (c) and $10^{-4}$ (d) KeV. 
 Green (Orange) regions correspond to normal (inverted) ordering of the light neutrino mass spectrum and cover all possible configurations for the CP phases. The gray (blue) regions are  experimentally excluded  by $0\nu 2 \beta$ experiments (end-point tritium beta decay experiments) while the dashed lines correspond to the future sensitivities for KATRIN (blue) and neutrinoless double beta decay (gray), see Table~\ref{tab:futuresensitivities}.}
  \label{fig:3+1:0nbb}
\end{figure}

The effect of this new contribution can be seen in Fig.~\ref{fig:3+1:0nbb}. 
In (a) we show the standard picture for the three active Majorana neutrino case (${\rm SM}_\nu$), with the colored bands covering the possible variation of the CP phases.
In the other three panels, we display the results after adding a fourth neutrino with increasing impact on the effective Majorana mass.
Notice that in the latter cases there is a new CP phase related to $U_{e4}$, which may affect the size of the predicted bands for both normal and inverted ordering cases.

The first important information inferred from Fig.~\ref{fig:3+1:0nbb} is that the presence of a sterile neutrino can strongly impact the prediction for $0\nu 2 \beta$, leading to different and possibly augmented ranges  for the effective mass $m_{ee}$,  when compared to the SM predictions, and this for both normal and inverted ordering of the light neutrino mass spectrum. This can be seen  by comparing  Fig.~\ref{fig:3+1:0nbb}(a) and Fig.~\ref{fig:3+1:0nbb}(b) for the NO case.  These changes depend of course on the parameter space for $m_4$ and $U_{e4}$, more precisely on the combination $m_4|U_{e4}|^2$. 

As can be seen in these plots, the picture changes when the sterile neutrino is compatible with a kink in the beta energy spectrum according to Eq.~(\ref{KATRINsensitivity}).
 For instance, when $m_4|U_{e4}|^2=10^{-4}$~KeV, one cannot distinguish the NO from the IO regions, see Fig.~\ref{fig:3+1:0nbb}(d). 
 Interestingly, there are cases where an observation of a signal in neutrinoless double beta decay (assuming a severe control on the nuclear matrix elements uncertainties) will not necessarily imply an inverted ordering of the light neutrino mass spectrum even for very small $m_{\rm lightest}$, compare for instance Fig.~\ref{fig:3+1:0nbb}(c) and Fig.~\ref{fig:3+1:0nbb}(a). 
 Alternatively, a non-observation in future $0\nu 2 \beta$ experiments would not rule out the IO, contrary to the ${\rm SM}_\nu$ case, if $m_4 |U_{e4}|^2$ is large enough (since there could be a cancellation as can be seen in   Fig.~\ref{fig:3+1:0nbb}(d)). 
There is also the possibility that $m_4|U_{e4}|^2$ is  smaller (even if $m_4$ is in the KeV mass region)  such that the fourth neutrino contribution is of the same order as the ${\rm SM}_\nu$ one, and this can even lead to a strong cancellation in $m_{ee}$ for particular choices of the CP violating phases. 

\begin{figure}[t!]
 \begin{center}
\begin{tabular}{cc}
  \includegraphics[width=0.49\textwidth]{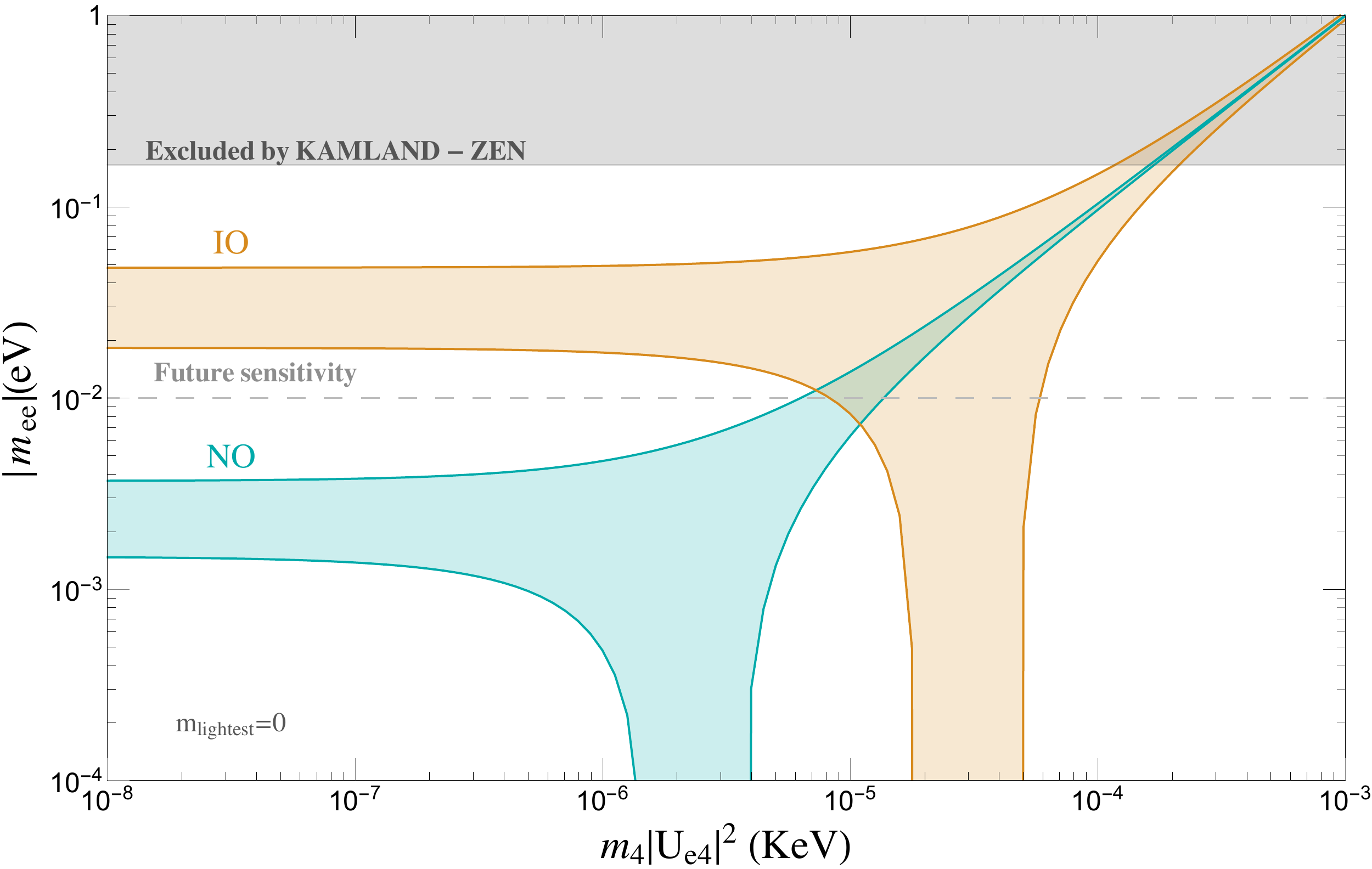}  &
   \includegraphics[width=0.49\textwidth]{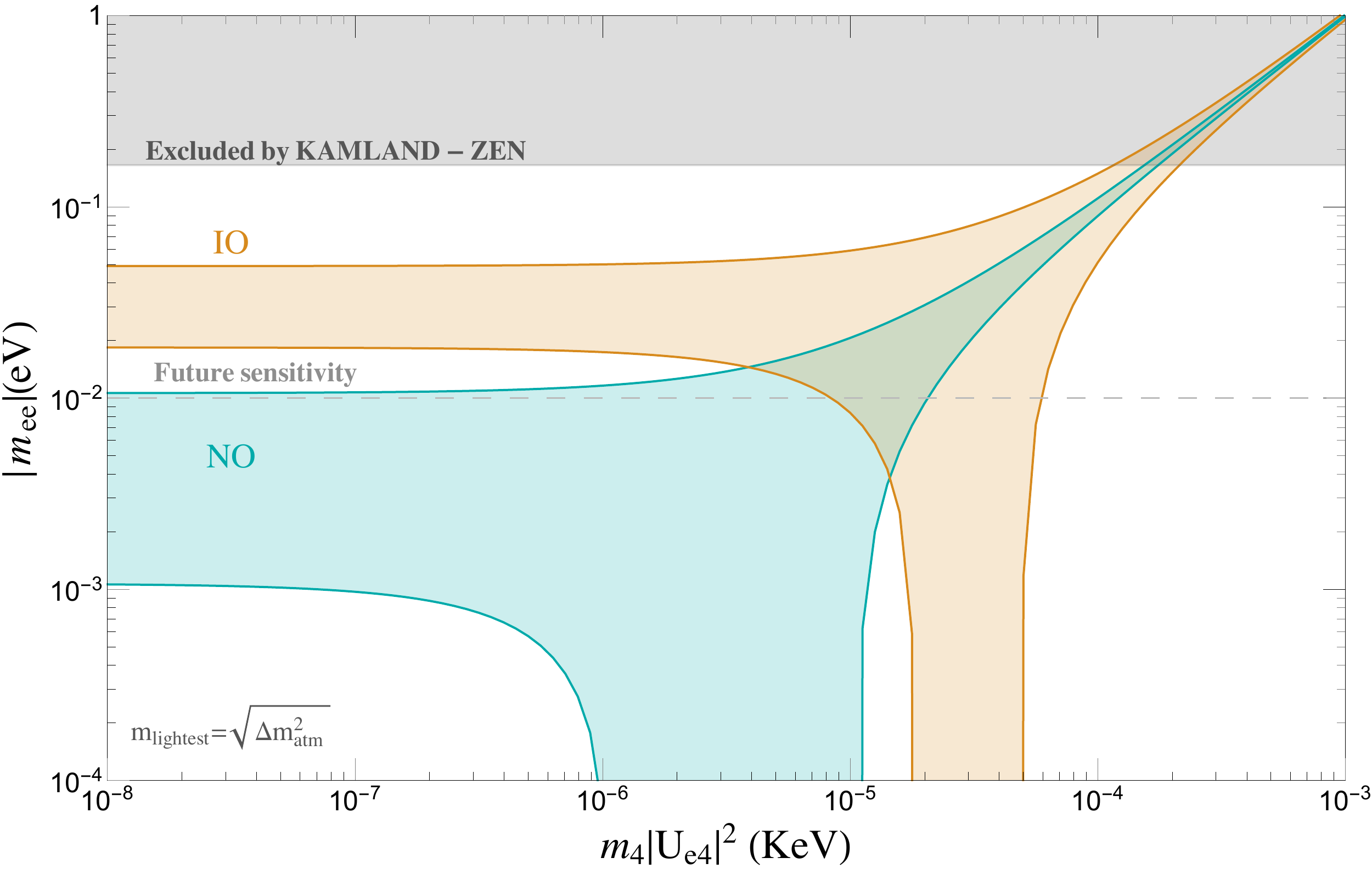}\\
   (a)&(b)
\end{tabular}
\end{center}
 \caption{Majorana effective mass $|m_{ee}|$ in  3+1 model as a function of  $m_4|U_{e4}|^2$ for two representative values for the mass of  the lightest neutrino $m_{\text{lightest}}=0$ (a), and  set to the atmospheric mass scale, $m_{\text{lightest}}=\sqrt{\Delta m^2_{\text{atm}}}$ (b). Lines and color codes as in Fig. \ref{fig:3+1:0nbb}.}
 \label{fig:3+1:m4Ue4}
  \end{figure}

We refrain from showing the several situations but instead address this cancellation when we explore the synergy between an observation of a  kink in  the KATRIN energy spectrum and a signal in $0\nu 2 \beta$. This is displayed in Fig.~\ref{fig:3+1:m4Ue4} where instead we show $|m_{ee}|$ as a function of $m_4|U_{e4}|^2$ for representative values of $m_{\text{lightest}}$. 
We show the particular case where $m_{\text{lightest}}=0$, Fig.\ref{fig:3+1:m4Ue4}(a) - anticipating the discussion for the seesaw models we consider in this study as their minimality imposes that the lightest (active) neutrino is massless, as discussed in Section \ref{sec:model} - and  the case in which the lightest neutrino mass is about the atmospheric oscillation scale, $\sqrt{\Delta m^2_{\text{atm}}}$, Fig.~\ref{fig:3+1:m4Ue4}(b). Interestingly, the latter case almost corresponds to the lower bound that future $0\nu 2 \beta$ experiments could probe in the normal ordering case. 
Nevertheless, the general behavior is the same in both panels: $|m_{ee}|$ is ${\rm SM}_\nu$-like for small values of $m_4|U_{e4}|^2$, while it is dominated by the sterile neutrino for large values of $m_4|U_{e4}|^2$. 
In the transition between the two regimes, when both active and sterile contributions are comparable, $|m_{ee}|$ may suffer the above mentioned cancellations, although the critical value of $m_4|U_{e4}|^2$ is very dependent on the ordering and the value for $m_{\rm lightest}$.

It is important to remark that Fig.~\ref{fig:3+1:m4Ue4} is valid  for any $m_4\ll125$~MeV, including of course the KeV neutrino we are interested in. 
In the case of KATRIN sensitivity, Eq.~(\ref{KATRINsensitivity}), this region would correspond to $m_4|U_{e4}|^2>10^{-6}$~KeV.
If  indeed KATRIN confirms the presence of a kink, then one could  draw a vertical line on Figs. \ref{fig:3+1:m4Ue4} (a) or (b) and infer a prediction for $|m_{ee}|$.
This prediction would correspond to an interval for $|m_{ee}|$ whose size would depend on the value of $m_4|U_{e4}|^2$.
The extreme case would be that this line lies in the cancellation regions where the contribution  $m_4 U_{e4} ^2$ cancels exactly $m_{ee}^{({\rm SM}_\nu)}$, and this could happen for several possible combinations of the CP violating phases. 
We can also see that the right part of the plots is in tension with present upper bounds on $|m_{ee}|$. 
Therefore, if KATRIN observes a kink with  $m_4 |U_{e4}|^2\gtrsim 3\times10^{-4}$~KeV, a more involved model than the 3+1 would be required.

The analysis can be extended to the case of the $3+N$ model where all sterile neutrino masses lie below the nuclear threshold ($m_i\ll 125$ MeV, $i=4\dots N$), or where the effect of the heavier neutrinos on $m_{ee}$ is negligible\footnote{This could happen, for instance, if the neutrinos are much heavier than $p^2$, their mixing to the electron very small, or if they form (pseudo-)Dirac pairs, as in the LSS or ISS models.}. 
In such case, our discussion for the 3+1 case can be easily generalized by replacing the role of $m_4|U_{e4}|^2$ by an effective heavy mass $|m_{ee}^{\rm heavy}|$ given by 
\begin{equation}
m_{ee}^{\rm heavy}= 
 \sum_{i=4}^{N}m_i U_{ei}^2\,.
\end{equation}
We have explicitly conducted the analysis for  the case $N=2$, since it accounts for the minimal amount of sterile neutrinos needed for generating light neutrino masses in a Type-I seesaw mechanism.  
Then one cannot draw direct predictions for $m_{ee}$ in the case where  KATRIN sees a kink, since it gives information on only one sterile state $m_4 |U_{e4}| ^2$ and not on the sum $|\sum_{i=4}^{N}m_i U_{ei}^2|$, unless KATRIN observes a second kink.
If the two kinks are well separated in mass, this could correspond to a Type-I seesaw where the two sterile neutrinos are in the mass range $[1, 18.5]$~KeV. 
On the other hand, if the two kinks are close in mass, they could point towards an approximate lepton number conserving scenario with quasi-degenerate sterile neutrinos. 
We refer to the latter as a {\it double-kink} signature.
This leads us to consider the (minimal) seesaw models we have presented in Section \ref{sec:model}, where one could potentially generate a neutrino spectrum such that the heavy states are in KATRIN's regime, while agreeing with neutrino data (as well as the several constraints discussed in  Appendix~\ref{sec:constraints}).

%%%%%%%%%%%%%%%%%%%%%%%%%%%%%%%%%%%%%%%%%%%%%%%%%%%%%%%%%%%%%%

\subsection{Type-I seesaw with two RH neutrinos}
\label{secResultsTypeI}

\begin{figure}[t]
 
 \begin{center}
\begin{tabular}{cc}
  \includegraphics[width=0.49\textwidth]{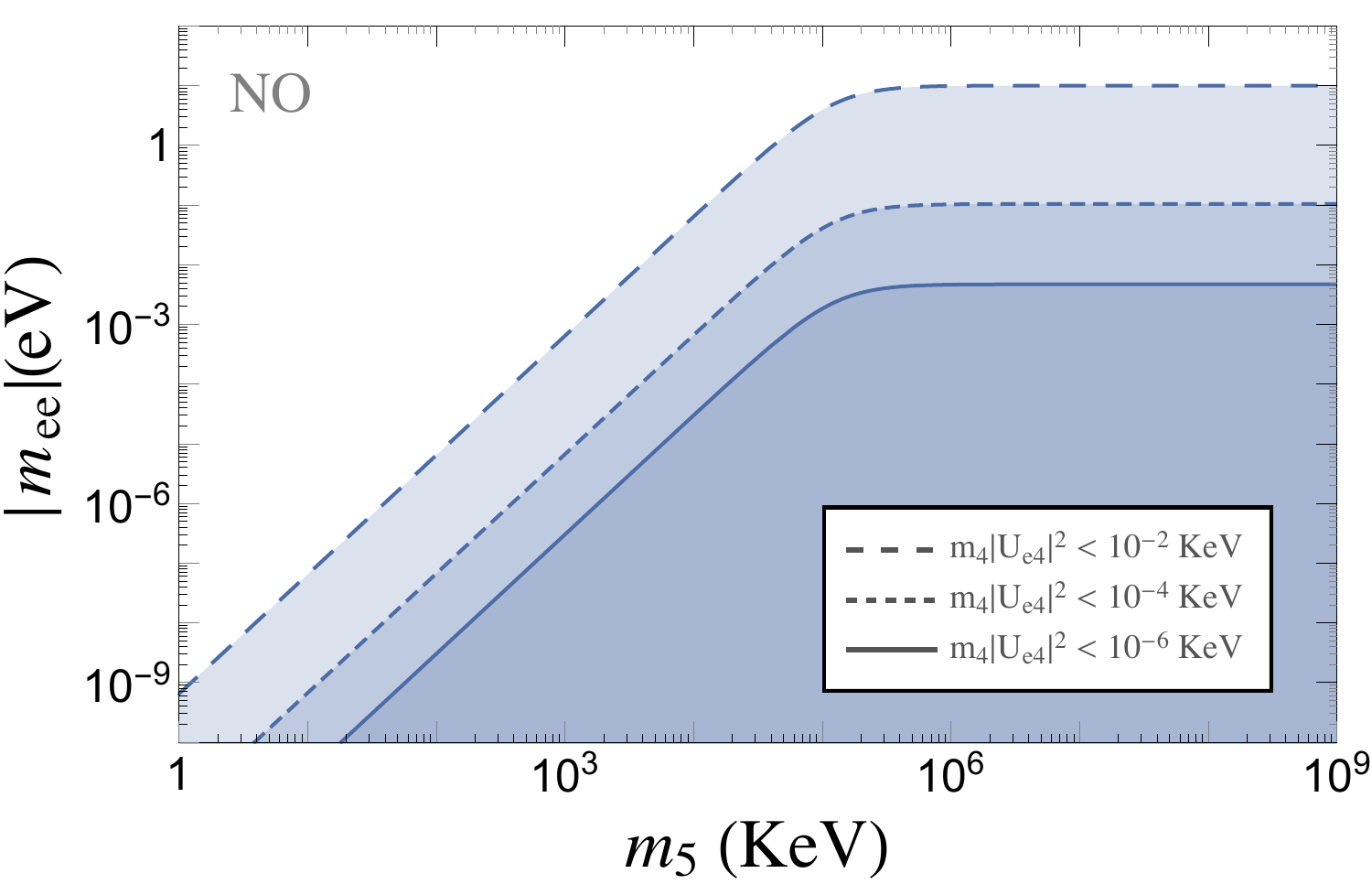}  &
   \includegraphics[width=0.49\textwidth]{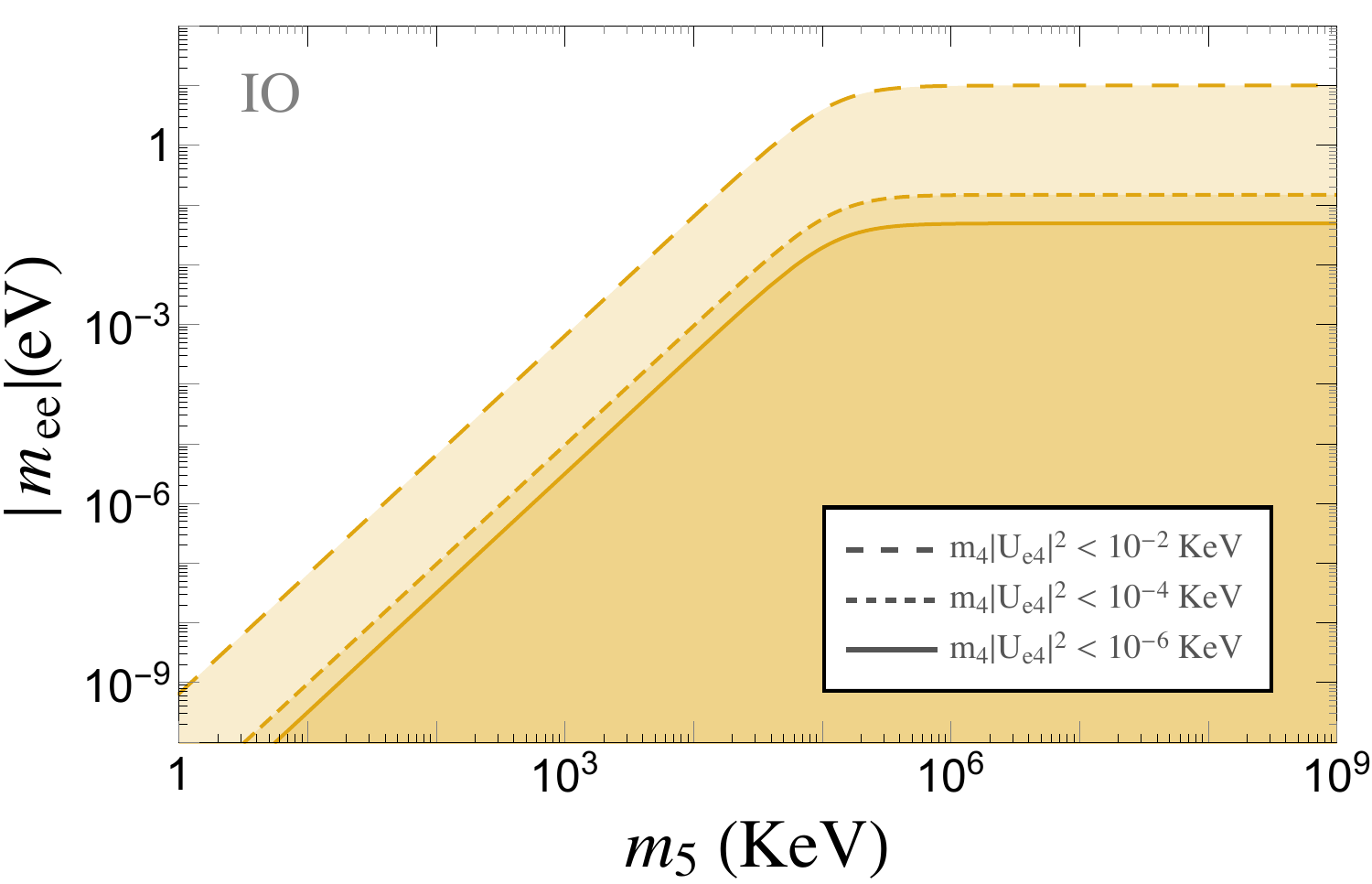}\\
   (a)&(b)
\end{tabular}
 \caption{Analytical prediction area for Majorana effective mass $m_{ee}$ in the Type-I seesaw with two RH neutrinos as a function of $m_5$, assuming a kink in the KATRIN beta spectrum for both ordering of the light neutrino spectrum NO (a) and IO (b). } \label{fig:gen-Type-1:0nbb-analytic}
 \end{center}

\end{figure}

We assume that one of the two RH neutrinos is within KATRIN sensitivity (kink in the beta spectrum) and consider the following three possible cases: when the second sterile neutrino mass is within KATRIN sensitivity as well (two kinks): $m_5\in [1, 18.5]~{\rm KeV}$; when it is above the tritium beta decay threshold energy, but below the nuclear double beta decay Fermi momentum, $m_5\in [18.5~{\rm KeV}, 125\ {\rm MeV}]$, and finally when $m_5\gg 125$~MeV.

\begin{figure}[tp]
\begin{center}
\begin{tabular}{cc}
 \includegraphics[width=0.49\textwidth]{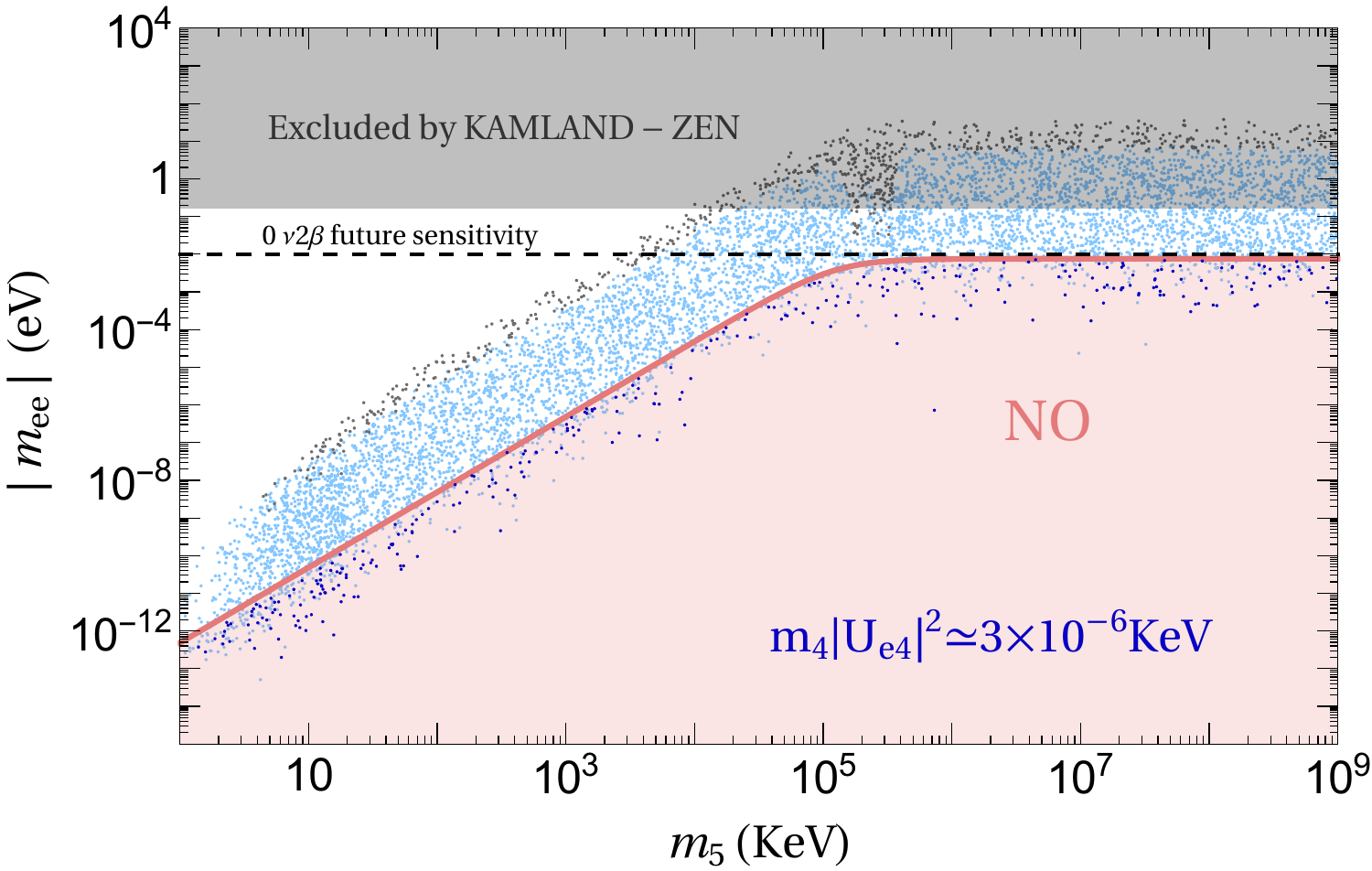}   &   
  \includegraphics[width=0.49\textwidth]{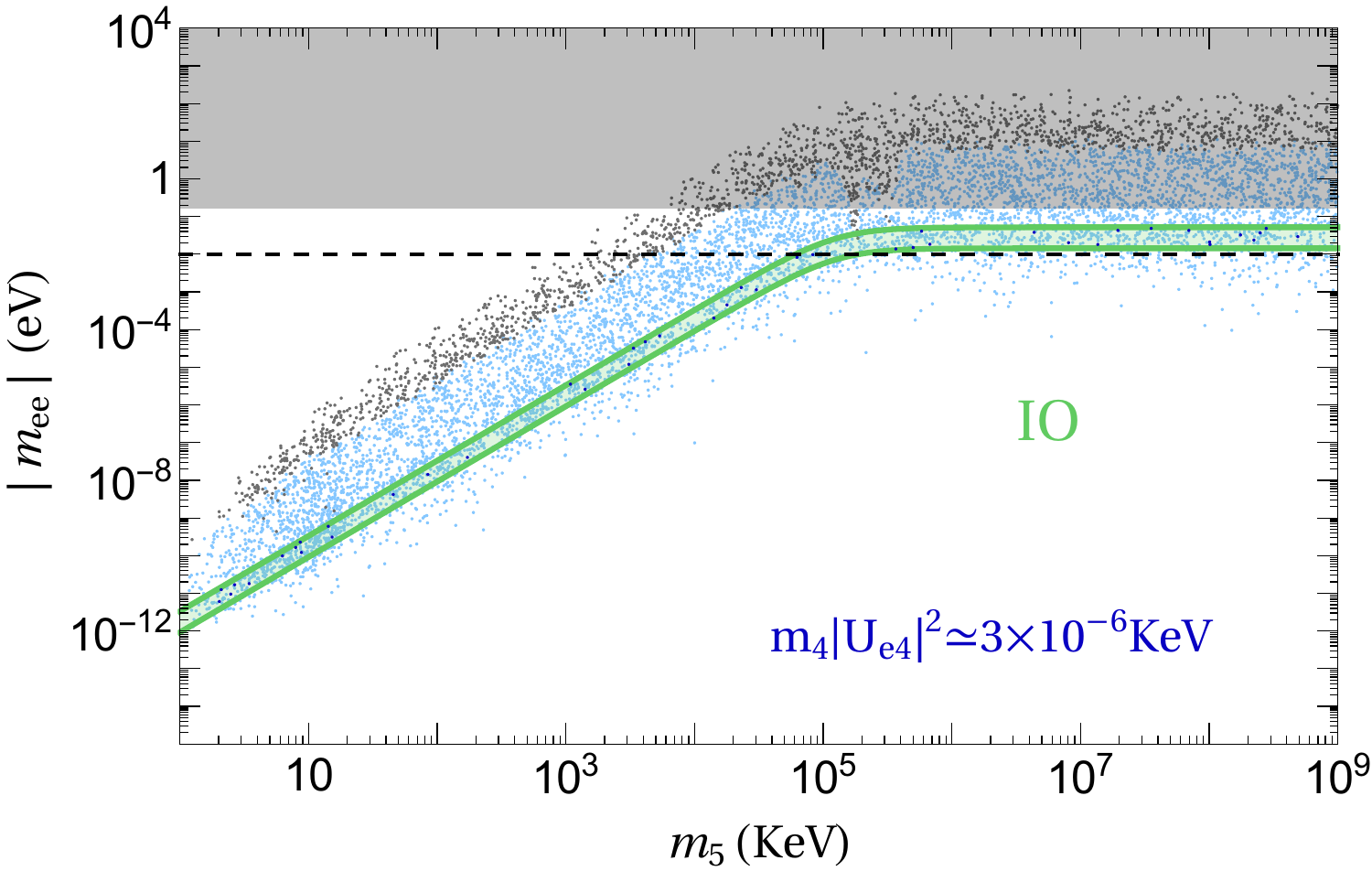}
 \\(a)&(b)\\
  \includegraphics[width=0.49\textwidth]{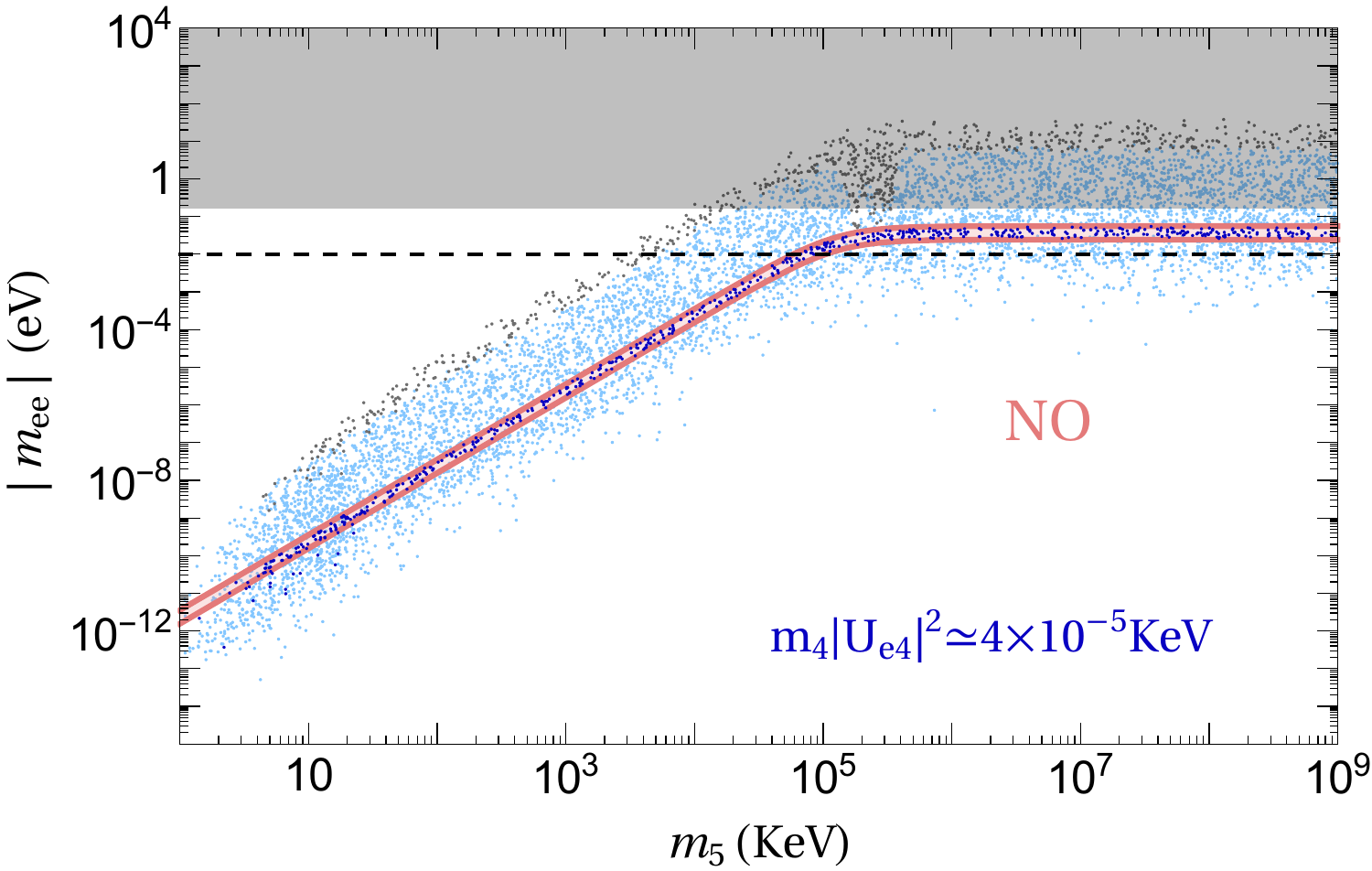}  
 & 
    \includegraphics[width=0.49\textwidth]{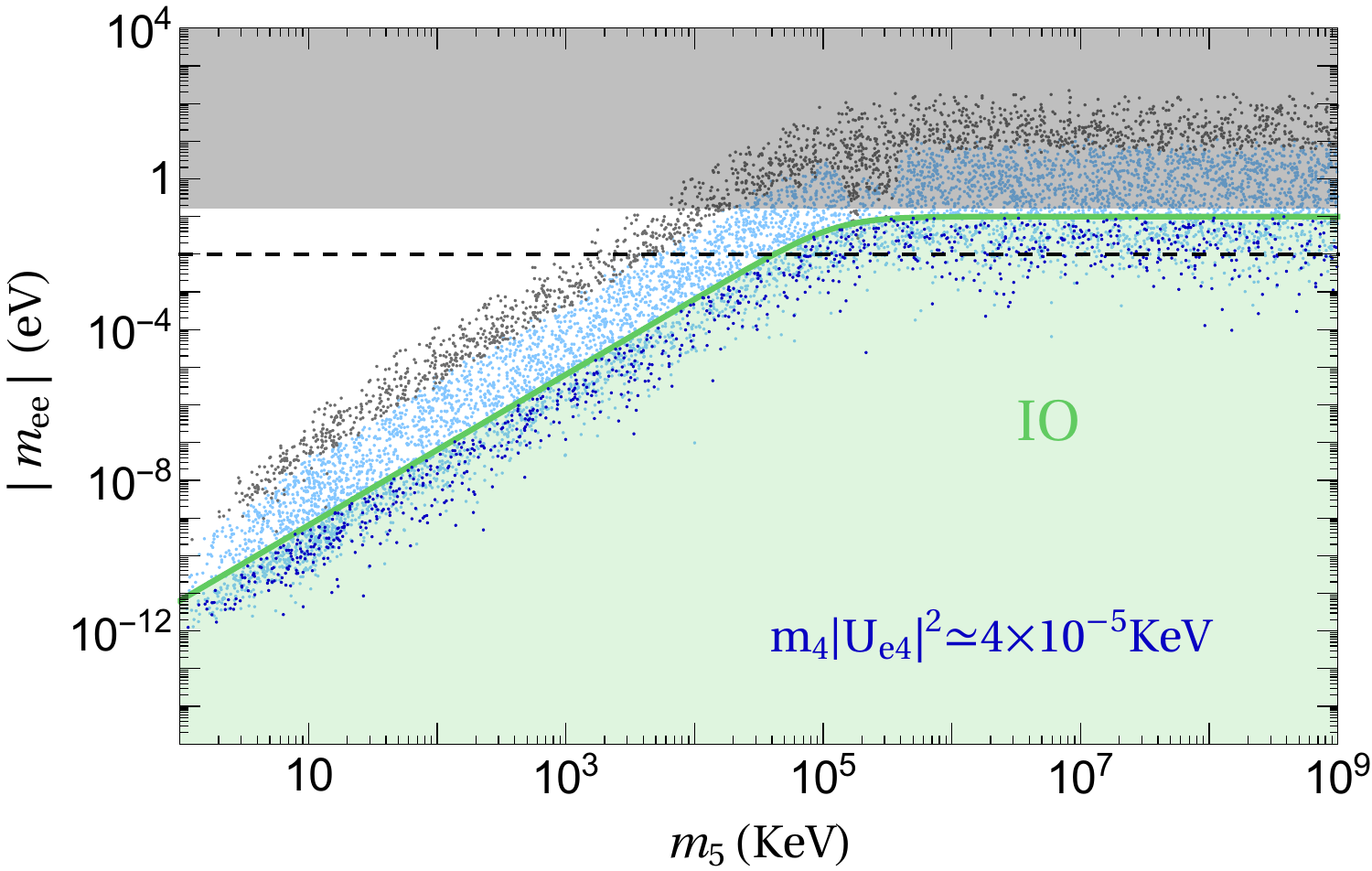}  \\(c)&(d)
    \\
  \includegraphics[width=0.49\textwidth]{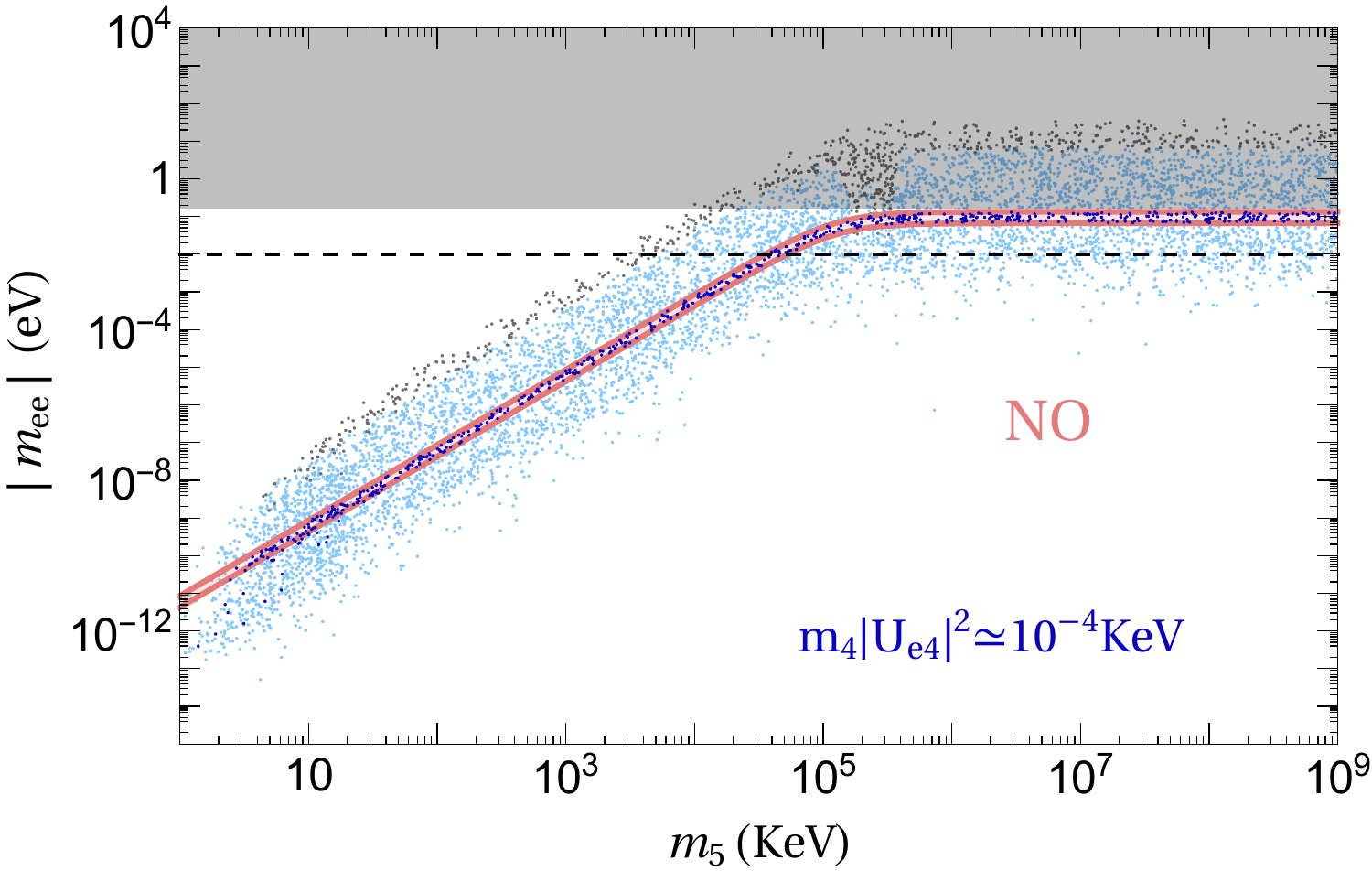}  
 & 
    \includegraphics[width=0.49\textwidth]{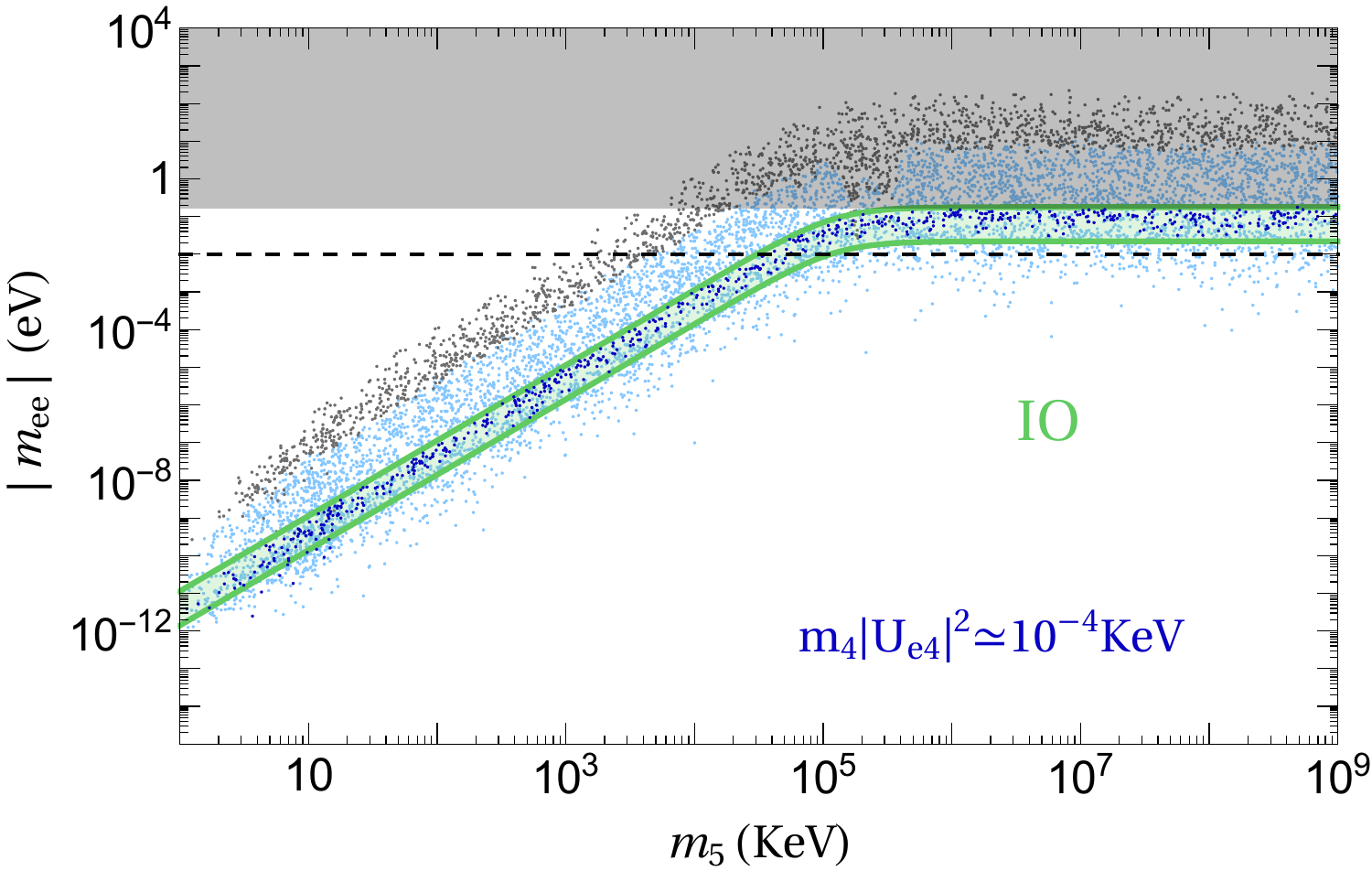}  \\(e)&(f)\end{tabular}
\end{center}
 \caption{Majorana effective mass $|m_{ee}|$ as a function of the heaviest sterile neutrino mass in the Type-I seesaw with two RH neutrinos in the situation where the lightest sterile neutrino is compatible with a kink in KATRIN beta spectrum for both NO (left) and IO (right) of the light neutrino mass spectrum. Light blue points are solutions compatible with neutrino data and the several constraints. Gray points are those not complying with at least one constraint. The gray  regions are  experimentally excluded  by $0\nu 2 \beta$ experiments while the dashed lines correspond to the future sensitivity of $0\nu 2 \beta$ experiments, Table \ref{tab:futuresensitivities}.
On each panel, the red (NO) and green (IO)  areas correspond  to analytical prediction for $|m_{ee}|$ when we allow a 30\% deviation from a  chosen central value:  $m_4|U_{e4}|^2\simeq 3\times10^{-6}$ (top), $4\times10^{-5}$(middle), $10^{-4}$(bottom) KeV. The solutions contained within these areas are highlighted in dark-blue.}
  \label{fig:Type1-band}
\end{figure}

Following the seesaw condition Eq.~(\ref{eq:seesaw.condition}) and the discussion thereafter, we expect that the $0\nu 2 \beta$ effective mass will always be below the analytical upper limit shown in Fig.~\ref{fig:gen-Type-1:0nbb-analytic}, where $|m_{ee}|$ is represented as a function of $m_5$ for different  ranges for $m_4|U_{e4}|^2$ (chosen so that the fourth neutrino state is compatible with a kink in KATRIN beta spectrum).  Depending on the values of all the remaining parameters (mixings and CP violating phases), the mass of the second sterile neutrino can be in the three above mentioned regimes. The cancellation in $|m_{ee}|$ for light values of $m_5$ below the nuclear threshold,  as well as the saturation line when $m_5$ is very large (above the $\sim$ TeV scale) is clearly manifest in both panels of Fig.~\ref{fig:gen-Type-1:0nbb-analytic} corresponding to normal ordering (a)  and inverted ordering (b), thus following the prediction of Eq.~(\ref{eq:meeseesaw}).
This cancellation is stronger when $m_5<18.5$~KeV, i.e., in the case of two possible kinks in the KATRIN beta spectrum.

 One can also notice on this figure how the predictions for $|m_{ee}|$ evolve depending on the light neutrino mass spectrum ordering and when the position of the kink in KATRIN (value of $m_4$) changes, in agreement with Eq.~(\ref{eq:meeseesaw}). 
For instance, when $m_4|U_{e4}|^2< 10^{-6}$ KeV, the allowed (analytical) region for $m_{ee}$ as well as the corresponding maximal prediction is higher in the IO case than in the NO one, although in both cases the predictions are close to the experimental future sensitivity reach, see Table \ref{tab:futuresensitivities}. 
When the hypothetical vertical line in Fig.~\ref{fig:3+1:m4Ue4},  that would correspond to an observation of a kink in KATRIN, moves from left to the right, the predictions for $|m_{ee}|$ become less  sensitive to the ordering of the light neutrino spectrum. Equivalently, this would correspond  to  increasing the maximal predictions for $|m_{ee}|$ until one could not distinguish between the normal and the inverted ordering cases.

In order to have an estimate for the predictions for beta and neutrinoless double beta decays for the Type-I seesaw model, we use the parametrization  given in Eq.~(\ref{paramtypeI}). 
With the hypothesis of at least one kink in KATRIN beta spectrum, the results obtained after having scanned over all the parameter space are presented in Fig.~\ref{fig:Type1-band} showing  $|m_{ee}|$ as a function of the heaviest sterile neutrino mass $m_5$, for both NO (left) and IO (right) orderings of the light neutrino mass spectrum. 
In all panels of Fig.~\ref{fig:Type1-band}, blue points correspond to the solutions compatible with neutrino data and the phenomenological bounds, while the gray points are those not complying with at least one constraint (most of them are not compatible with direct searches constraints). 
On each panel, the red (NO) and green (IO)  areas correspond  to analytical prediction for $|m_{ee}|$ when we allow a 30\% deviation from a  chosen central value of  $m_4|U_{e4}|^2\simeq 3\times10^{-6}$~KeV (top), $4\times10^{-5}$~KeV (middle) and $10^{-4}$~KeV (bottom). The numerical solutions contained within these areas are highlighted in dark-blue. 
At first sight, one can confirm that the solutions compatible with neutrino data and the phenomenological bounds (blue points) are always within the analytical area discussed after Fig.~\ref{fig:gen-Type-1:0nbb-analytic},  confirming  the analytical expectation of Eq.~(\ref{eq:meeseesaw}).  

 One can also  see that when $m_4|U_{e4}|^2\simeq 3\times10^{-6}$ KeV, there is no lower analytical bound in the NO case, see  Fig.~\ref{fig:Type1-band}(a), due to possible cancellations in $|m_{ee}|$, as can be easily confirmed from Fig.~\ref{fig:3+1:m4Ue4}(a), however  the maximal values for $|m_{ee}|$ in this case  are close  to future sensitivity reach. On the other hand, the situation is different  in the IO case, see Fig.~\ref{fig:Type1-band}(b), as for this value for $m_4|U_{e4}|^2$, the analytical region (green) is very narrow; one could thus infer the value of $m_5$ (as well as information on the remaining Type-I seesaw parameters) based on the interplay between both observables (possible kink in KATRIN and a signal in $0\nu 2 \beta$).
 The role of the two orderings is reversed for $m_4|U_{e4}|^2\simeq 4\times10^{-5}$ KeV, Fig.~\ref{fig:Type1-band}(c) and (d). 
 Nevertheless, these cancellations do not occur in some other cases for  $m_4|U_{e4}|^2$ and we can have a very narrow (thus predictive) bands for both ordering cases,  as one can see  in Fig.~\ref{fig:Type1-band}(e) and (f).
 
\begin{figure}[t!] 
  \centering
  \begin{tabular}{cc}
  \includegraphics[width=.48\textwidth]{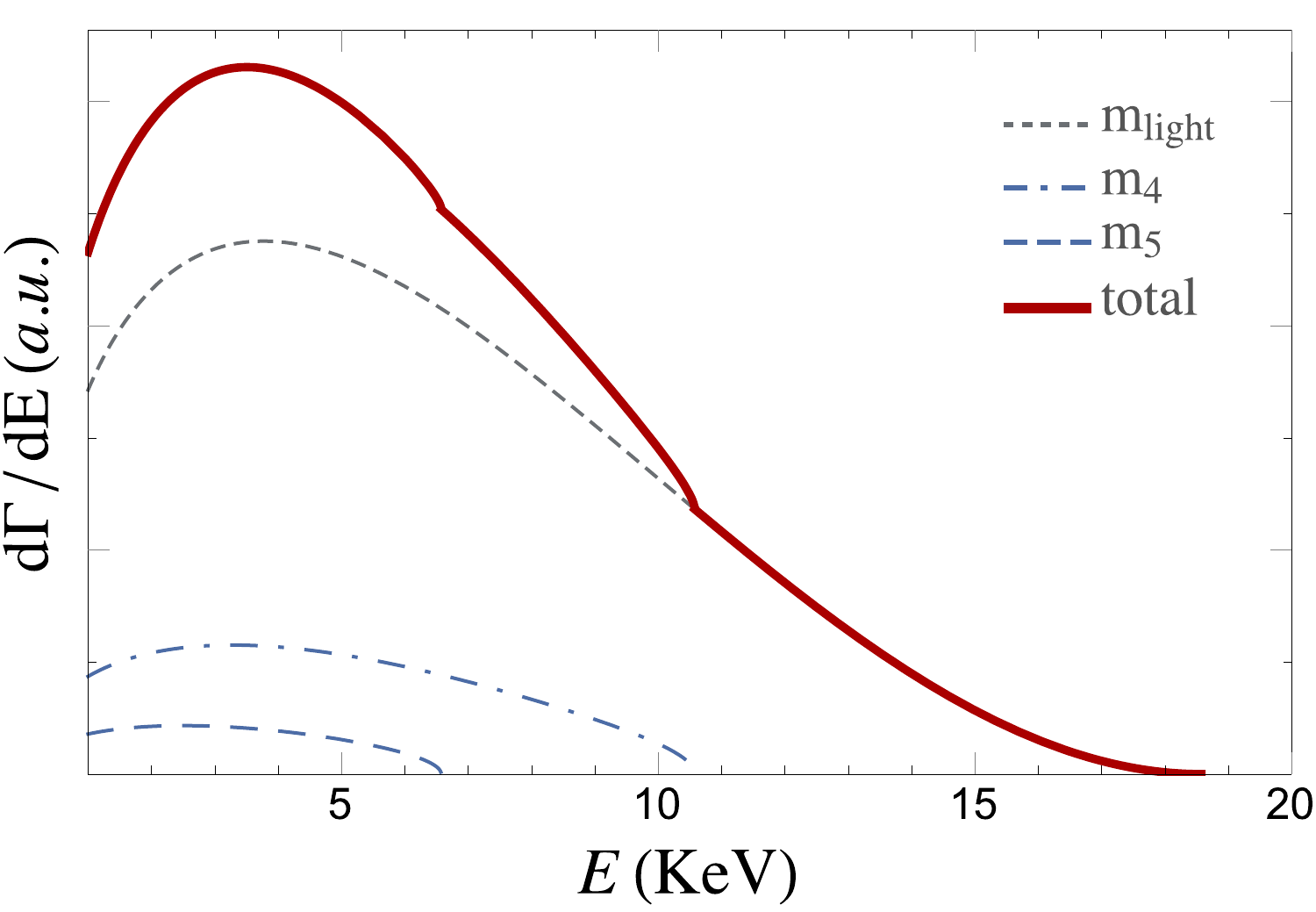}&  \includegraphics[width=.5\textwidth]{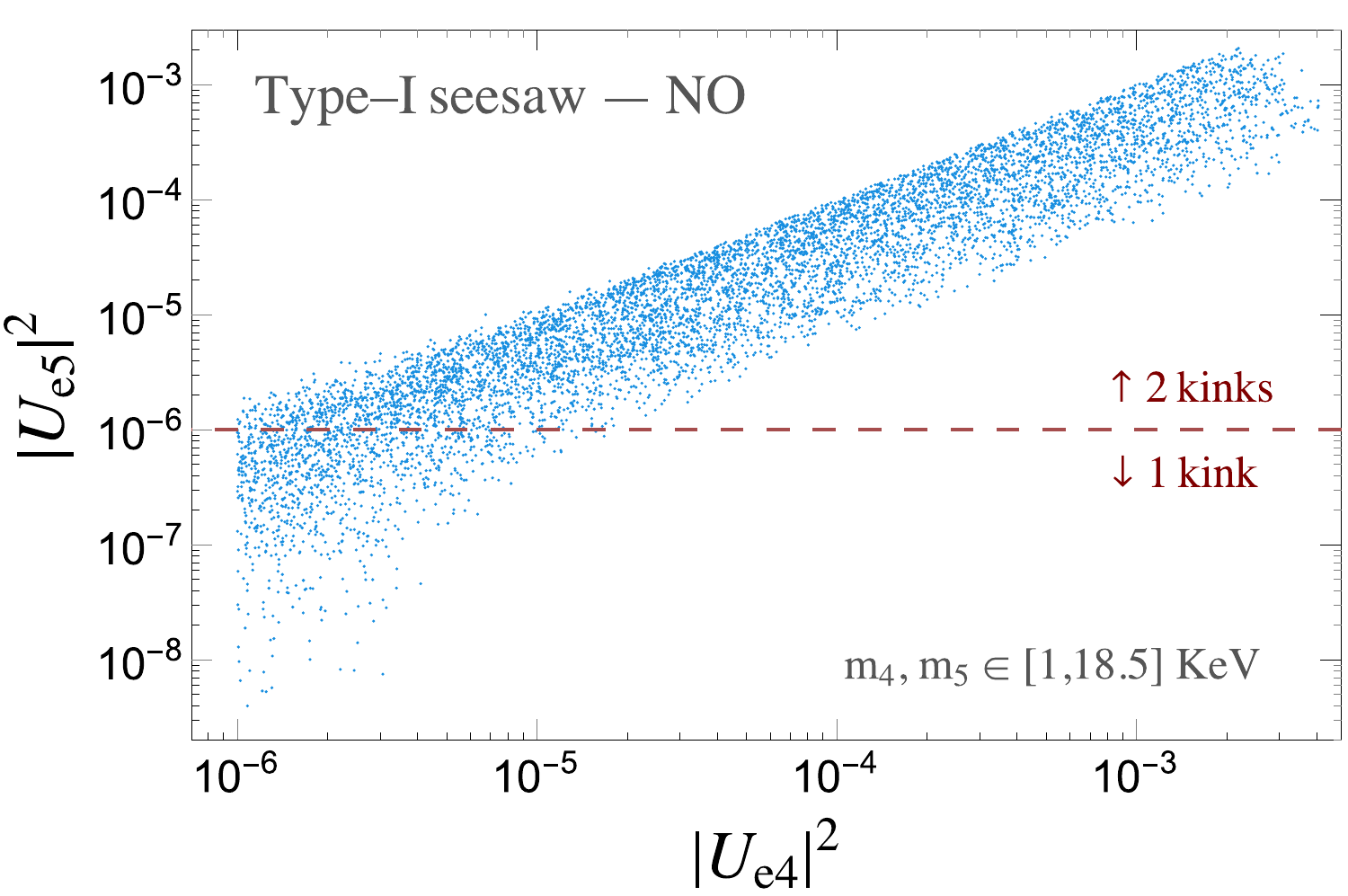}\\
(a)&(b)\end{tabular}
 	\caption{(a):  Tritium beta spectrum in the presence of two kinks. For this example, we set $m_4=8$~KeV, $m_5=12$~KeV, $|U_{e4}|^2=0.2$ and $|U_{e5}|^2=0.1$. 
(b): Active-sterile mixings in the case where the two sterile neutrinos of the Type-I seesaw are within KATRIN mass regime. The horizontal dashed red  line corresponds to the assumed (conservative) sensitivity of KATRIN.}
\label{fig:onetwokinks}
\end{figure}

 Thus, with the help of the above examples, one could discuss the interplay between both observables, the tritium beta decay energy spectrum and neutrinoless double beta decay effective mass. 
For example, we found in this analysis  favorable cases where one could have a signal in KATRIN (a kink) and a signal in  $0\nu 2 \beta$ experiments, pointing toward a Type-I seesaw with one sterile state below $18.5$~KeV and a heavier one  with a mass below the nuclear threshold of $\sim 125$~MeV. 
This situation would correspond for instance to the dark-blue points contained in the bands of Fig.~\ref{fig:Type1-band}(b), (c), (e) and (f), some of them giving also prediction for the light neutrino mass ordering. 
On the other hand, a non-observation of $0\nu 2 \beta$ would imply an upper bound on $m_5$, see Fig.~\ref{fig:Type1-band}(b), (c) (e) or (f), and therefore one would expect to detect the second sterile neutrino in a low energy experiment. 
An interesting case of the latter situation would be when the second sterile neutrino has a mass $m_5$ lying also in the KeV regime, since KATRIN could potentially observe it as a second kink, as shown in Fig.~\ref{fig:onetwokinks}(a).

Finally, we explore  the case where both sterile neutrino masses are below $18.5$~KeV, thus no signal in $0\nu 2 \beta$ is expected whatever is the ordering of the light neutrino mass spectrum.
As already said, KATRIN could then signal the presence of one kink or two kinks, depending on the remaining physical parameters. This is exemplified in Fig.~\ref{fig:onetwokinks}(b), where the scatter plot shows viable solutions  imposing the lightest sterile neutrino within KATRIN sensitivity and complying with neutrino data (for NO in this case) and the relevant constraints (blue points).
As can be seen, depending on the  active-sterile mixing $|U_{e5}|^2$, there are viable solutions compatible with the presence of a second kink in KATRIN spectrum, the situation of which is favored for large $|U_{e4}|$ mixings. 
Similar results where found for the inverted ordering case. 

Interestingly, among the solutions in Fig.~\ref{fig:onetwokinks}, there are some points  where the positions of the two kinks are very near  (double-kink),  implying that the sterile neutrinos are very close in mass. 
This would be the situation one would find in the case of a scenario with approximate lepton number symmetry as the LISS model.

\subsection{Results for the LISS scenario}
\label{sec:LISSresults}

\begin{figure}[t!]
  \centering
  \begin{tabular}{cc}
  \includegraphics[width=.47\textwidth]{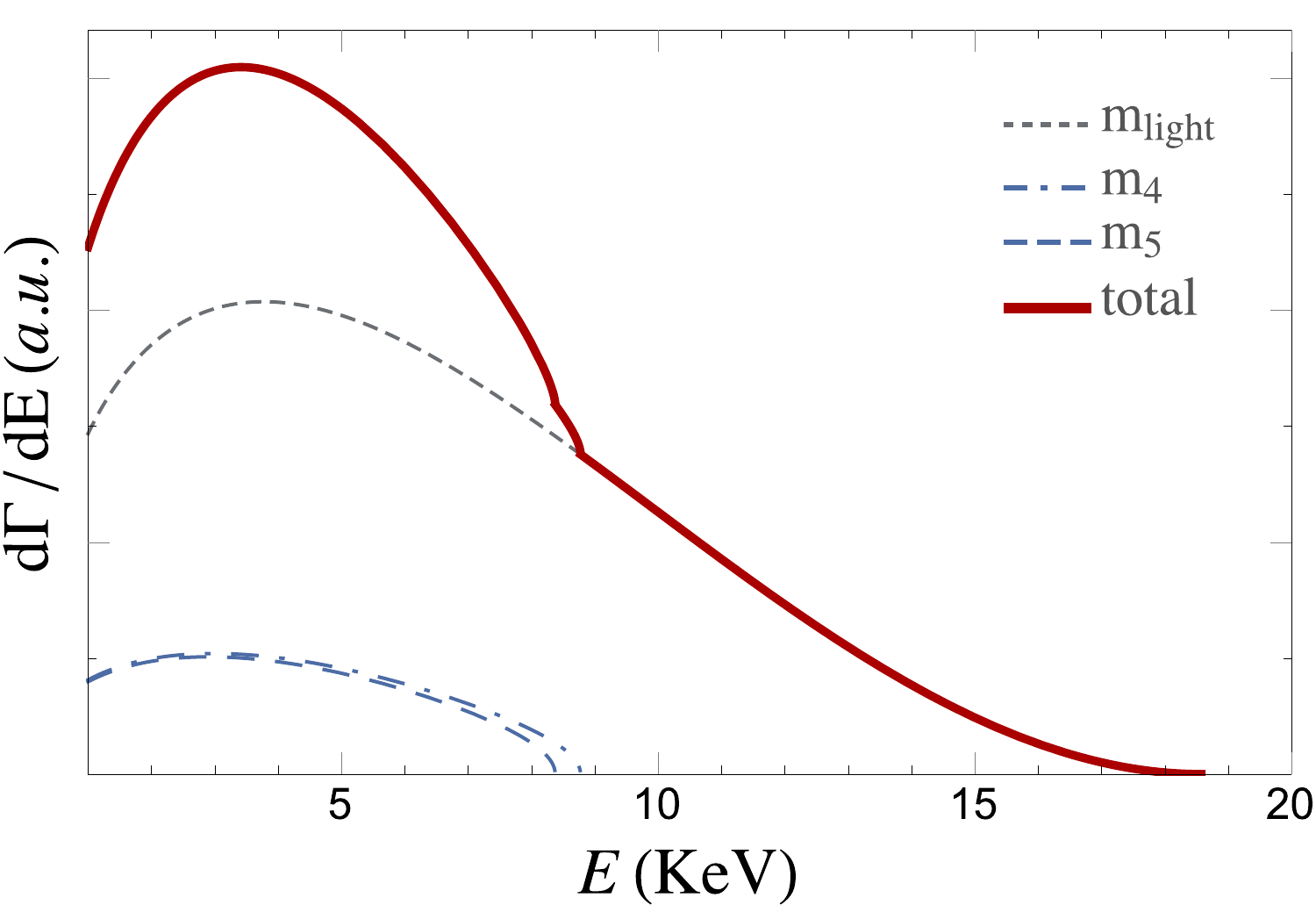}&  \includegraphics[width=.51\textwidth]{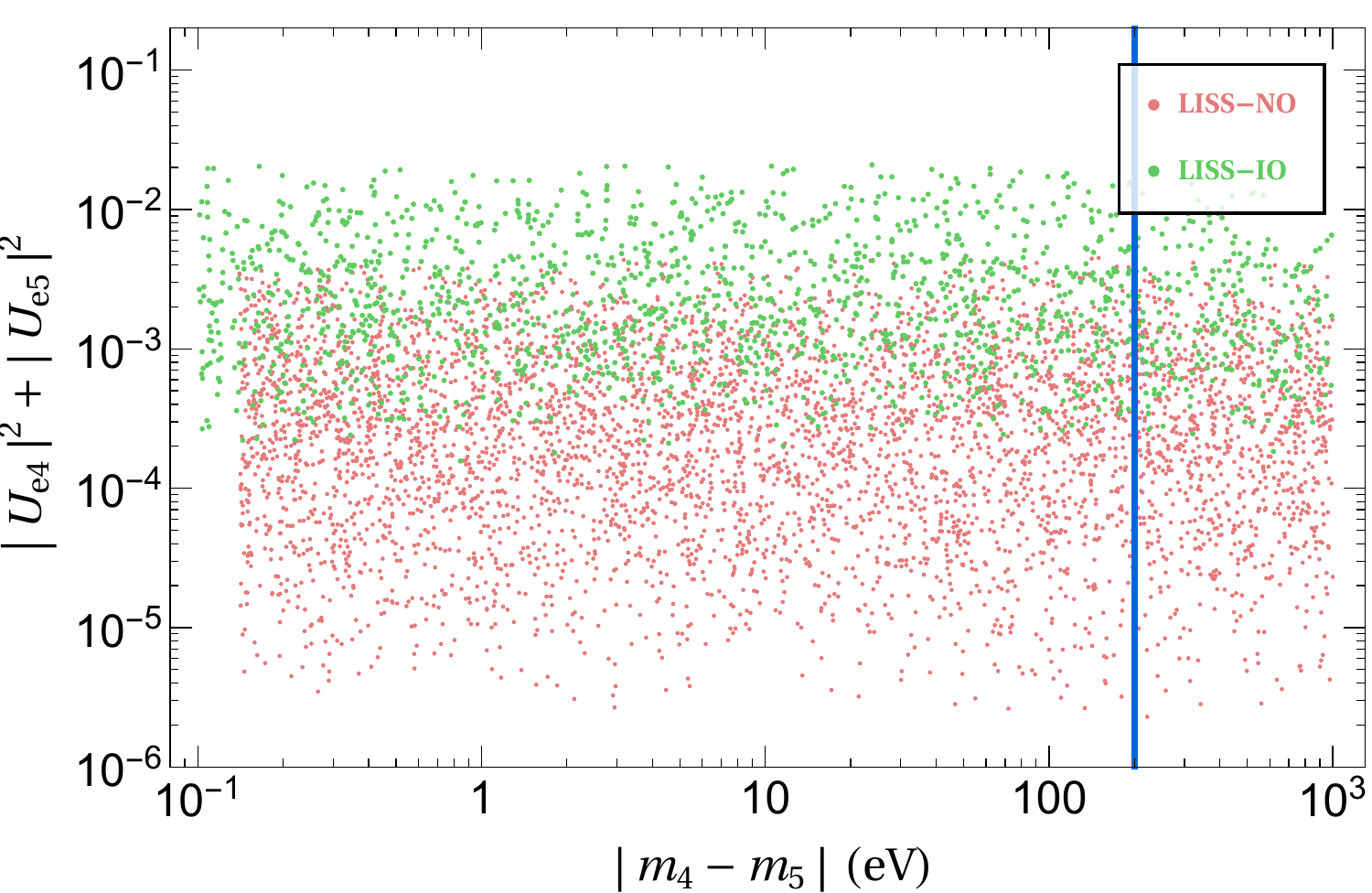}\\
(a)&(b)\end{tabular}

  \caption{(a): Tritium beta spectrum in the presence of a double-kink. For this example, we set $m_4=9.8$~KeV, $m_5=10.2$~KeV and $|U_{e4}|^2=|U_{e5}|^2=0.2$. 
 (b): Active-sterile mixings $|U_{e4}|^2+|U_{e5}|^2$ versus the mass splitting $|m_4-m_5|$ where the two sterile neutrinos of the LISS model are within KATRIN mass regime and complying with all the relevant constraints. The vertical blue  line corresponds to the assumed KATRIN resolution for a double-kink.}  \label{fig:LNC:split}	
\end{figure}

As discussed in Section \ref{sec:model}, in the LISS scenario the heavy (mostly) sterile neutrinos are close in mass and their mixings are similar in size, $|U_{e4}|\simeq |U_{e5}|$, while the deviations from this degeneracy are controlled by the LNV parameters. 
If their masses are  below $18.5$ KeV, one could expect the presence of a double-kink in KATRIN energy spectrum, as shown in Fig.~\ref{fig:LNC:split}(a). Whether KATRIN would be able to resolve a double-kink depends on its energy resolution and on the LNV parameters defining the LISS model, thus on the mass splitting between the two sterile states. 

In the case of the presence of a kink in the beta decay energy spectrum and no positive signal in $0\nu 2 \beta$, nor in other low energy experiment, KATRIN would help exploring this model by studying in detail the observed kink. 
If the mass splitting between  the two sterile neutrinos is below KATRIN's energy resolution\footnote{For the discussion in this section, we will use 200~eV as a benchmark, although a dedicated study would be needed.}, the experimental signature would be a single kink with a {\it size} of $ |U_{e4}|^2 + |U_{e5}|^2$.
On the other hand, if the resolution is high enough to resolve the double-kink, KATRIN would be able to provide information on the mass splitting and, therefore, on the LNV parameters of the model. 

In Fig.~\ref{fig:LNC:split}(b) we show  viable solutions complying with neutrino data and constraints for  NO (red) and IO (green) orderings of the light neutrino spectrum. 
We choose  $|U_{e4}|^2+|U_{e5}|^2$  for the y-axis as it corresponds to the height of a kink due to the presence of the two sterile states in the  case where  KATRIN would not be able to resolve the double-kink. 
Notice that they are in general larger in the IO case than in the NO one, therefore assuming KATRIN observes indeed a kink with a very large value for $|U_{e4}|^2+|U_{e5}|^2$, this would favor the inverted mass ordering of the light neutrino spectrum.
The x-axis represents the mass splitting $|m_4-m_5|$, with a vertical blue line showing the considered KATRIN energy resolution for  this discussion. 
We decided to stop at 1~KeV when LNV is not well justified, since then the mass splitting is comparable to the sterile neutrino masses and one should study the Type-I seesaw as in section~\ref{secResultsTypeI}.
From this figure we see that the LISS model predicts a broad range of mass splittings. 
If the points are to the right of the blue line, KATRIN would be able to resolve the double-kink and measure the mass splitting. 
On the other hand, if they are to the left, the resolution would not be enough to resolve it and it would be observed as a single kink, which would  nevertheless allow KATRIN to set upper limits on the mass splitting. 
Consequently, in both cases KATRIN  would provide information about the sources of  LNV in this scenario.

This study can be enlarged to other low-scale seesaw frameworks like the $\nu$MSM model~\cite{Asaka:2005pn,Asaka:2005an} where a Type-I seesaw is at work with three RH neutrinos 
whose mass spectrum and couplings to the active
states are severely constrained by the requirements of having a successful BAU and providing a viable dark matter candidate. 
The $\nu$MSM-predictions for  $0\nu 2 \beta$ Majorana effective mass has been addressed in for instance~\cite{Bezrukov:2005mx,Asaka:2011pb} and more recently in~\cite{Asaka:2016zib} ($\nu$MSM with 3 RH neutrinos) and in~\cite{Hernandez:2016kel} (Type-I seesaw with 2 RH neutrinos). It has been shown that  the prediction for $|m_{ee}|$ can be sizable ($\sim 140$~meV at max) if the two heaviest RH neutrinos have a mass close to the nuclear momentum $\simeq 200$~MeV with a large mass difference ($\simeq 1$~MeV) in the case of IO, and this  only when the CP phases and the mixings are appropriately aligned. 
The lightest RH neutrino (in the case of the $\nu$MSM with 3 RH neutrinos) being in the KeV mass region in order to provide a viable dark matter candidate can in principle impact KATRIN's energy spectrum, however, due to the smallness of its mixing to the electron neutrino, it  is beyond the sensitivity of KATRIN.

\section{Conclusions}\label{sec:conclusion}
In this study we have  explored the viability of minimal extensions of the Standard Model involving sterile neutrinos (namely  the $3+N$ model and low-scale seesaw mechanisms with two sterile neutrinos) and study their possible impact in both neutrinoless double beta decay neutrino effective Majorana mass and in the KATRIN tritium beta decay energy spectrum.

In our numerical analysis, we explore different mass regimes for the extra fermions, the active-sterile mixings as well as the different CP violating phases. In particular, we identify and discuss the regimes where it is possible to have (at least) one KeV neutrino within the sensitivity of KATRIN and the other one much heavier giving rise to a possible signal in $0\nu 2 \beta$ experiments, for both orderings of the light neutrino spectrum. 

In the Type-I seesaw, assuming that one of the two RH neutrinos is within KATRIN's sensitivity 
(a kink in the energy spectrum), we addressed the three following possible cases: \begin{itemize}
\item[i)] when the second RH neutrino mass is within KATRIN sensitivity as well ($m_5\in [1, 18.5]~{\rm KeV}$), KATRIN has the potential to detect a second kink, while the $0\nu 2 \beta$  effective Majorana mass vanishes; 
\item[ii)] when it is above the tritium beta decay threshold energy, but below the nuclear double beta decay Fermi momentum ($m_5\in [18.5~{\rm KeV}, 125\ {\rm MeV}]$), then there is not such 
a second kink and the Majorana effective mass still vanishes;
\item[iii)] when $m_5\gg 125$~MeV, 
one can expect to observe a signal in $0\nu 2 \beta$  experiments. 
\end{itemize}
Moreover,  in the first case, the two kinks could be close in mass such that KATRIN could observe a double-kink in the energy spectrum, pointing towards a Type-I seesaw extended with an input, for instance related to an approximate lepton number conservation. We have explored this  possibility by studying a model combining the linear and the inverse seesaw mechanisms with two sterile neutrinos (LISS).

In summary, our study shows how the interplay between the two observables, the electron energy spectrum in KATRIN tritium $\beta$ decay  and  the $0\nu 2 \beta$  effective Majorana mass,  can help constraining the sterile neutrino parameter space and ultimately  discriminating  between motivated low-scale seesaw realizations involving at least one KeV sterile neutrino. 

%%%%%%%%%%%%%%%%%%%%%%%%%%%%%%%%%%%

\section*{Acknowledgments}
The authors would like to thank S.~Mertens for
fruitful discussions. They are also thankful to C.~Garcia-Garcia for her help and suggestions. 
We acknowledge partial support from the European Union Horizon 2020
research and innovation programme under the Marie Sk{\l}odowska-Curie: RISE
InvisiblesPlus (grant agreement No 690575)  and 
the ITN Elusives (grant agreement No 674896). 
%%%%%%%%%%%%%%%%%%%%%%%%%%%%%%%%%%%

\appendix

%%%%%%%%%%%%%%%%%%%%%%%%%%%%%%%%%%%%%%
\section{Constraints on sterile fermion hypothesis}\label{sec:constraints}

The modifications of the vertices in Eq.~(\ref{eq:lag}) due to the presence of the (rectangular $3\times (3+N)$) leptonic mixing matrix imply deviations from unitarity  of the ($3\times 3$) PMNS mixing matrix; moreover  having
massive  sterile neutrinos as final decay products  can possibly induce further 
deviations from the SM theoretical expectations. Consequently, scenarios with sterile fermions are severely constrained and any extension of the SM involving these states must comply with neutrino data and with several constrains,  some of them being stringent. 
This Appendix collects the most stringent constraints on the SM extensions we considered, providing those relevant for the regimes we explore.
%%%%%%%%%%%%%%%
\subsection{Neutrino oscillation data}
Any of the extensions we consider in this work has to comply with neutrino oscillation parameters (squared neutrino mass differences and their corresponding mixings). The recent fit from neutrino data  give the following ranges for mixing angles and masses, which corresponds to $3\sigma$ confidence level~\cite{Gonzalez-Garcia:2014bfa,Esteban:2016qun},
\begin{eqnarray}
0.272\leq\sin^2\theta_{12}\leq0.346,\quad
0.418\leq\sin^2\theta_{23}\leq0.613,\quad
0.01981\leq\sin^2\theta_{13}\leq0.02436,\\
6.80\leq\frac{\Delta m_{21}^2}{10^{-5}~\mathrm{eV}^2}\leq8.02,\quad
2.399\leq\frac{\Delta m_{31}^2}{10^{-3}~\mathrm{eV}^2}\leq2.593	,
\hspace{2cm}
\end{eqnarray}
in the case of normal ordering of the light neutrino spectrum, and 
\begin{eqnarray}
0.272\leq\sin^2\theta_{12}\leq0.346,\quad
0.435\leq\sin^2\theta_{23}\leq0.616,\quad
0.02006\leq\sin^2\theta_{13}\leq0.02452,\\
6.80\leq\frac{\Delta m_{21}^2}{10^{-5}~\mathrm{eV}^2}\leq8.02,\quad
-2.562\leq\frac{\Delta m_{32}^2}{10^{-3}~\mathrm{eV}^2}\leq-2.369,
\hspace{1.7cm}
\end{eqnarray}
for the inverted ordering case. 
In our analysis we considered the best-fit central values given above, allowing for a deviation of $\sim 5\%$.

%%%%%%%%%%%%%%%
\subsection{Direct searches}
We have used in our analysis 
the most recent and up-to-date available constraints ~\cite{Atre:2009rg,Abada:2017jjx,Deppisch:2015qwa, Alekhin:2015byh} form direct searches on the parameter
space of the SM extended by additional massive Majorana fermions. For masses below 10~MeV, we have used Ref.~\cite{Atre:2009rg}, and for 
masses between 10~MeV and 100~GeV, we have used the constraints discussed in~\cite{Abada:2017jjx} (and references therein).

%%%%%%%%%%%%%
\subsection{Cosmological and astrophysical constraints}
Cosmological observations, see for instance Ref.~\cite{Smirnov:2006bu,Kusenko:2009up,Adhikari:2016bei,Abazajian:2017tcc}, 
 put severe constraints on sterile neutrinos with a mass below the GeV scale as they can constitute an important fraction of  dark matter  impacting structure formation, which is constrained
by Large Scale Structure and Lyman-$\alpha$ data.
In addition, their mixings to the active neutrinos may induce the radiative decays $\nu_i \to \nu_j
\gamma$ that  are well constrained by cosmic X-ray searches, see for instance Refs.~\cite{Loewenstein:2008yi,Loewenstein:2012px}. There are also severe  constraints from Big Bang Nucleosynthesis (BBN) and/or CMB, which would be relevant in the very low mass regime for sterile neutrinos. Indeed, sterile neutrinos decay and in order to evade BBN and CMB constrains, they have to decay before the onset of BBN otherwise they can modify quantities as the primordial helium abundance~\cite{Kawasaki:2000en} and the effective number of neutrinos $N_{\rm eff}$ as well as produce effects in structure formation.
On top of that, a large mixing of the KeV neutrinos with the active ones would also imply an overabundance of dark matter. A dedicated study of  the specific extension of the SM with two sterile neutrinos has been conducted in for instance~\cite{Hernandez:2014fha} where all possible cases for the (light) masses (and active-sterile mixings) regimes of the two extra neutrinos  have been considered, showing that the models are strongly constrained for masses below ${\mathcal{O}}(100)$ MeV. 
All these constraints are put assuming a standard cosmology  and can be evaded if a
non-standard cosmology is considered~\cite{Gelmini:2004ah,Gelmini:2008fq}, or when sterile
neutrinos couple to  a dark sector~\cite{Bezrukov:2017ike,Dasgupta:2013zpn} or in an extended Left-Right symmetric sector, allowing to evade bound from X-rays~\cite{Barry:2014ika}.  For this reason, these constraints are not  taken into account in our numerical analysis. There are other alternatives to the seesaw mechanism discussed above, as for instance in models with an extended Higgs sector, and the possibility that the Majorana mass term for neutrinos is  generated by the expectation value of a gauge-singlet Higgs boson. This is for instance considered in~\cite{Petraki:2007gq} in which  the relic abundance of sterile neutrinos does not necessarily depend on their mixing angles, allowing  the free-streaming length to be smaller than in the case when the dark matter is produced by neutrino oscillations (Dodelson-Widrow mechanism)~\cite{Dodelson:1993je}, relaxing thus the bounds from X-rays. %
\vspace{0.5cm}

\noindent

\section{parametrization of the lepton mixing matrix  for the 3+2 model}\label{sec:para3+2}
%\subsubsection{}
The mixing matrix $U$ for the $N=2$ can be parametrized as 
\begin{equation}
 U=R_{45}R_{35}R_{25}R_{15}R_{34}R_{24}R_{14}R_{23}R_{13}R_{12}
~\mathrm{diag}\left(1,e^{i\varphi_2},e^{i\varphi_3},e^{i\varphi_4},e^{i\varphi_5}\right),
\label{eq:mixing}
\end{equation} 
where $R_{ij}$ is the rotation matrix between $i$ and $j$. 
For instance, the rotation matrix $R_{45}$ is explicitly given by 
\begin{equation}
R_{45}=\left(
\begin{array}{ccccc}
1 & 0 & 0 & 0 & 0\\
0 & 1 & 0 & 0 & 0\\
0 & 0 & 1 & 0 & 0\\
0 & 0 & 0 & \cos\theta_{45} & \sin\theta_{45}e^{-i\delta_{45}}\\
0 & 0 & 0 & -\sin\theta_{45}e^{i\delta_{45}} & \cos\theta_{45}\\
\end{array}
\right),
\end{equation}
and likewise for the other matrices $R_{ij}$ (in terms of $\theta_{ij}$ and  $\delta_{ij}$.).

Since the number of Dirac phases is 6 for the case where $N=2$, four Dirac phases
$\delta_{ij}$ can be eliminated: we thus set $\delta_{12}=\delta_{23}=\delta_{24}=\delta_{45}=0$. 

Notice that the mixing matrix in Eq.~(\ref{eq:3+1rot}) for the case  $N=1$ can be obtained by taking the $4\times4$
sub-matrix after substituting $R_{i5}=\1$ (and putting $\varphi_5=0$) in Eq.~(\ref{eq:mixing}).

%%%%%%%%%%%%%%%%%%%%%%%%%%%%%%%%%%%

\end{document}